\documentclass{ws-rv9x6}
\usepackage{seceqn}
\newcommand{\eq}[1]{Eq.~(\ref{#1})}
\newcommand{\fig}[1]{Fig.~\ref{#1}}
\newcommand{\dotv}{\mbox{\boldmath\(\cdot\)}}
\newcommand{\cross}{\mbox{\boldmath\(\times\)} }

\newcommand{\sgn}{{\mathrm{sgn}\,}}
\renewcommand{\Re}{{\rm Re}}
\renewcommand{\Im}{{\rm Im}}

\newcommand{\kvec}{{\mathbf{k}}}

\newcommand{\vvec}{{\mathbf{v}}}
\newcommand{\xvec}{{\mathbf{x}}}

\newcommand{\khat}{\hat{\kvec}}
\newcommand{\xhat}{\hat{\xvec}}
\newcommand{\kDe}{k_{\rm De}}
\newcommand{\kDi}{k_{\rm Di}}
\newcommand{\kDs}{k_{\rm Ds}}
\newcommand{\ms}{m_{\rm s}}
\newcommand{\ompi}{\omega_{\rm pi}}
\newcommand{\omps}{\omega_{\rm ps}}

\newcommand{\Ts}{T_{\rm s}}
\newcommand{\Id}{{\sf\bf{I}}}

\begin{document}

\setcounter{chapter}{0}

\chapter{THE SCREENED FIELD OF A TEST PARTICLE}

\markboth{R.L. Dewar}{The Screened Field of a Test Particle}

\author{Robert L. Dewar}

\address{
Research School of Physics \& Engineering\\
The Australian National University\\
Cenberra ACT 0200, Australia\\
E-mail: robert.dewar@anu.edu.au}

\begin{abstract}
The screened field (forward field and wake) of a test particle moving at constant velocity through an unmagnetized
collisionless plasma is calculated analytically and numerically. This paper is based on unpublished material from my MSc thesis, supervised by the late Dr K.~C. Hines.
\end{abstract}

\section{Introduction}\label{sec:1_Intro}

Interest in the kinetic theory of interacting charged particles at University of Melbourne developed from the work of Dr Ken  Hines\cite{Hines_1955} in the 1950s---improving on a calculation by Landau\cite{Landau_1944} and using a formalism developed by Fano,\cite{Fano_1953} he calculated the slowing down distribution function of a charged ``test particle'' passing through a thin layer of material. The plasma theory group developing around Ken attracted a number of research students, including myself. The test particle diffusion problem provided a focus for a reading group on statistical physics\cite{Chandrasekhar_1943} and plasma kinetic theory,\cite{Balescu_1963,Montgomery_Tidman_1964} and research on the problem developed in several ways, including a relativistic generalization.\cite{Frankel_Hines_Dewar_1979}

The friction and diffusion coefficients of the Fokker--Planck equation suffer logarithmic divergences arising from the long-range nature of the Coulomb interaction. The fusion plasma theorists of the 1950s handled these divergences by the rather crude device of cutting off the interaction at the Debye length $1/k_{\rm D}$, the characteristic length occurring in the screened electrostatic potential, $q\exp(-k_{\rm D}r)/r$, in the neighbourhood of a static particle embedded in an ionized medium. While the logarithmic nature of the divergence makes the coefficients only weakly dependent on the cutoff,\cite{Frankel_Hines_Dewar_1979} the \emph{ad hoc} nature of this procedure was not very satisfactory from a fundamental point of view and this sparked research in the international theoretical physics community to find better approaches.  

In 1960, using very sophisticated formalisms, Balescu\cite{Balescu_1960a} and Lenard\cite{Lenard_1960} derived kinetic equations of the Landau form in which dynamical screening was incorporated through a frequency ($\omega$) and wavenumber ($\kvec$) dependent dielectric constant $\epsilon(\omega,\kvec)$.  The same Fokker--Planck--Landau equation (which is now known as the Balescu--Lenard equation) was derived by Thompson and Hubbard\cite{Thompson_Hubbard_1960,Hubbard_1961a,Hubbard_1961b} from simpler statistical physics arguments in which the diffusion coefficient was calculated from a fluctuation spectrum obtained by superimposing the dielectrically screened fields of independently moving particles.

This approach was called the \emph{dressed test particle picture} by Rostoker.\cite{Rostoker_1964} In this approach the unperturbed ``test particles'' replace the actual particles in the plasma. (In reality the trajectories are perturbed slightly by the fluctuations, giving rise to the linear dielectric response.)

The dressed test particle picture was developed in Fourier, $(\omega,\kvec)$, representation rather than in real space-time, $(\xvec,t)$, and thus it was difficult to visualize the actual nature of the screened potential surrounding each particle.  In Chapters 2 and 3 (unpublished) of my Master's Thesis,\cite{dewar67} which was supervised by Ken Hines, I calculated the screened potential in real space for a nonrelativistic plasma.\footnote{Chapter 1 contained a covariant relativistic plasma response function formalism that was later incorporated into a paper on energy-momentum tensors for dispersive electromagnetic waves.\cite{Dewar_77}} This contribution to the K.~C. Hines memorial volume is based on those chapters with only a few changes to make it self-contained and to improve readability.

For zero test particle velocity the solution is just the well-known
Debye potential $\exp(-k_{\rm D}r)/r$, but as the velocity is
increased we may expect qualitative changes to occur.  The work of
Pines and Bohm\cite{pines-bohm52} in which they considered forced
vibrations of the collective coordinates indicates that for a particle
moving slowly with respect to the electron thermal velocity
the Debye potential is distorted into a set of spheroidal
equipotentials centred on the particle and still decays exponentially
with distance.  We show that neither result is true, since the
criterion $k \lesssim k_{\rm D}$ is not an adequate criterion for the
existence of collective coordinates.  Similar results to ours were
found in this case by Rand\cite{rand59,rand60} by considering
individual particle trajectories in the self consistent field.  This
is simply a way of circumventing the use of the Vlasov equation but is
equivalent to it.  Rand makes an analogy between the symmetrical
result of Pines and Bohm and the Gibbs paradox of fluid dynamics, and
this analogy is upheld by the fact that
Majumdar\cite{majumdar60,majumdar63} and Cohen\cite{cohen61} get
results in agreement with Pines and Bohm by using fluid dynamical
treatments.

%
%

In this paper we consider a homogeneous isotropic magnetic-field-free
plasma.  We do not however restrict ourselves to a one-component
plasma.  In connection with the field around a small
satellite, considered by Kraus and Watson,\cite{kraus-watson58} it is necessary to consider the ions
since the satellite velocity is comparable more with the ion velocity
than the electron velocity.  In this plasma there are two modes of
longitudinal excitation, namely, ion acoustic waves and electron plasma waves,
but if the test particle is much slower than the electron thermal
velocity then only the ion waves can be excited---the case
considered by the above authors.  For ion waves not to be Landau damped
out of existence the electron Debye length must be much greater than
that of the ions, implying the electrons are much hotter than the
ions or that the ions have a much greater charge.  Kraus and
Watson\cite{kraus-watson58} also consider the case of a dense plasma,
in which local thermal equilibrium may be assumed.

Despite the preceding remarks about two-component plasmas the case of
infinite electron Debye length, which essentially reduces the problem
to a one component model, has received the bulk of our attention.
Most of our results therefore can only directly be compared with those
of Majumdar.\cite{majumdar60,majumdar63}  We show that his neglect of
Landau damping is not justified near the edge of the wake.  The
supersonic case receives brief attention in Sec.~\ref{sec:3ci_supersonicv0} and \ref{sec:3cii_supersonicv0}.

We note that Pappert\cite{pappert60} considered the effect of an
ambient magnetic field on the wake of a test particle, but we do not
treat this case.  A considerable body of Russian work on the details
of the satellite problem had also appeared by the time of this
thesis.\cite{gurevich64,panchenko-pitayevsky64,gurevich-pitayevsky64,gurevich-pitayevsky65}

\section{The Formal Solution \label{sec:2_Formal}}

The electrostatic potential $\varphi(\xvec,t)$ in the vicinity of a
test particle of charge $q$ moving rectilinearly at velocity $\vvec_0$
($v_0 \ll c$) through a homogeneous, stable dispersive dielectric medium
is obtained from standard linear response theory as
\begin{equation}
    \varphi = 
    \frac{q}{\varepsilon_0}\int\frac{d^3k}{(2\pi)^3}
    \frac{\exp[i\kvec\dotv(\xvec-\vvec_0 t)]}
         {k^2\epsilon(\kvec\dotv\vvec_0,\kvec)} \;.
    \label{eq:response}
\end{equation}
(Henceforth we consider the time to be $t = 0$, when the test particle
is at $\xvec = 0$, or, equivalently, represent the potential in a
frame moving with the test particle.)

\begin{figure}[th]		
	\centerline{\psfig{file=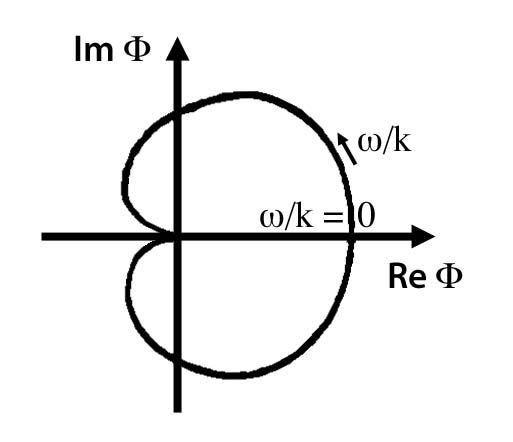,width=1.5in,angle=0}}
	\vspace*{1pt}
	\caption{The locus of $\Phi(\omega/k)$ in the complex plane as the 
	phase speed, $\omega/k$, is varied from $-\infty$ to $\infty$. The 
	point corresponding to $\omega/k = 0$ is the intersection of the locus with the $\Re \Phi$ axis as indicated.
	\label{fig:MScFig2_1}}
\end{figure}

In \eq{eq:response} $\epsilon(\kvec\dotv\vvec_0,\kvec)$ is the
\emph{dielectric constant}, which, for an isotropic, collisionless
unmagnetized plasma is given by
\begin{equation}
    \epsilon(\omega,\kvec) \equiv \epsilon(\omega,|\kvec|) =
    1 + \frac{\Phi(\omega/k)}{k^2} \;.
    \label{eq:dielectric}
\end{equation}
The function $\Phi$ is defined by
\begin{equation}
    \Phi(\omega/k) \equiv 
    \sum_s\omega^2_{{\rm p}s}
    \int_{-\infty}^{\infty} dv \frac{g_s'(v)}{\omega/k - v +i0} \;,
    \label{eq:Phidef}
\end{equation}
$\omega_{{\rm p}s}$ denoting the plasma frequency, $(e_s^2
n_s/\varepsilon_0 m_s)^{1/2}$ (SI units) for species $s$, with $n_s$ the
unperturbed number density, $m_s$ the mass, and $g_s(v)$
the one-dimensional projection of the unperturbed velocity
distribution function $f_s(\vvec)$. That is, in an arbitrary $x,y,z$ 
Cartesian coordinate system,
\begin{equation}
   g_s(v_z) \equiv 
    \frac{1}{n_s}\int\!\!\int dv_x dv_y f_s(\vvec) \;,
    \label{eq:gdef}
\end{equation}
the normalization factor $1/n_s$ being introduced so that $\int dv g_s
\equiv 1$.  Henceforth we take $s$ to denote electrons and a single
species of ion, denoted by subscripts e and i
respectively.

The assumed form of the ``hodograph'' of $\Phi$ for each species is
sketched in \fig{fig:MScFig2_1}, which is such as to ensure stability towards exponentially growing oscillations by the
Nyquist criterion [i.e. that the hodograph for $\epsilon(\omega,k)$ not
enclose the origin].

Equation~(\ref{eq:response}) becomes now
\begin{equation}		
    \varphi = 
    \frac{q}{\varepsilon_0}\int\frac{d^3k}{(2\pi)^3}
    \frac{\exp (i\kvec\dotv\xvec)}
	 {k^2+\Phi(\khat\dotv\vvec_0)} \;.
    \label{eq:screenedpotl}
\end{equation}

This is the starting point for all the following calculations.
We shall consider only test particle velocities much less than the
mean electron velocity so that we may approximate $\Phi_{\rm
e}(\omega/k)$ to the static value $k_{\rm De}^2 \equiv \Phi_{\rm
e}(0)$, the square of the (generalized) inverse Debye length for the
electrons.  Thus
\begin{equation}		
    \Phi(\omega/k) \approx 
    k^2_{\rm De} + \ompi^2
    \int_{-\infty}^{\infty} dv \frac{g_{\rm i}'(v)}{\omega/k - v +i0} \;.
    \label{eq:Phiapprox}
\end{equation}

Note that if the electrons are extremely hot with respect to the ions
then $k^2_{\rm De} \rightarrow 0$ and the ions form an essentially one
component plasma with the electrons forming a neutralizing background.
The physical explanation of this is that, although the electrons are
much lighter than the ions, they are moving too fast to be appreciably
deflected from their paths within the range of the test particle and
hence do not take part in the screening.  If, on the other hand, the
test particle has a velocity comparable with that of an average
electron then the \emph{ions} may be regarded as a uniform background
due to their inertia.  This case is therefore formally identical to
the case $k^2_{\rm De} = 0$.

We give finally the normalized Maxwellian and Lorentzian 
distribution functions for a species $s$ in a non-relativistic plasma together with the
corresponding polarization functions $\Phi_{\rm s}$.

\emph{Maxwellian case:} 
\begin{equation}
   g_{\rm s}(v) = 
    \left(\frac{\ms}{2\pi \Ts}\right)^{1/2}
    \exp -\frac{\ms v^2}{2\Ts}
    =
    \frac{\kDs}{(2\pi)^{1/2}\omps}
    \exp -\frac{1}{2}\left(\frac{\kDs v}{\omps}\right)^2
    \;,
    \label{eq:gMaxwell}
\end{equation}
where $\ms$ is the particle mass and $\Ts$ the temperature in energy 
units, giving
\begin{eqnarray}
    \Phi_{\rm s}\left(\frac{\omps}{\kDs}x\right) & = &
    \kDs^2
    \left[
       1
     - \sqrt{2}\,x\exp\left(-\frac{x^2}{2}\right)\Psi\left(\frac{x}{\sqrt{2}}\right)
     \right. \nonumber \\ & & \phantom{\kDs^2 [1\:} + \left.
     i\left(\frac{\pi}{2}\right)^{1/2}x\exp\left(-\frac{x^2}{2}\right) 
    \right] \;,
    \label{eq:Phi_Max}
\end{eqnarray}
for any dimensionless $x$, where $\kDs^2 \equiv e_{\rm s}^2n_{\rm
s}/\varepsilon_0 \Ts$ and
\begin{equation}
    \Psi(y) \equiv \int_{0}^{y}dt\;\exp t^2
    \; .
    \label{eq:Psidef}
\end{equation}


\emph{Lorentzian case:}
\begin{equation}
   g_{\rm s}(v) = 
    \frac{1}{\pi}\frac{u_{\rm s0}}{v^2 + u_{\rm s0}^2} \;.
    \label{eq:gLorentz}
\end{equation}
\begin{equation}
   \Phi_{\rm s}\left(\frac{\omps}{\kDs}x\right) = \frac{\kDs^2}{(1-ix)^2}
   \;,
   \label{eq:Phi_Lor}
\end{equation}
where $\kDs\equiv \omps/u_{\rm s0}$.

\section{Analytical approximations}\label{sec:3_Analapprox}

\begin{figure}[th]		
	\centerline{\psfig{file=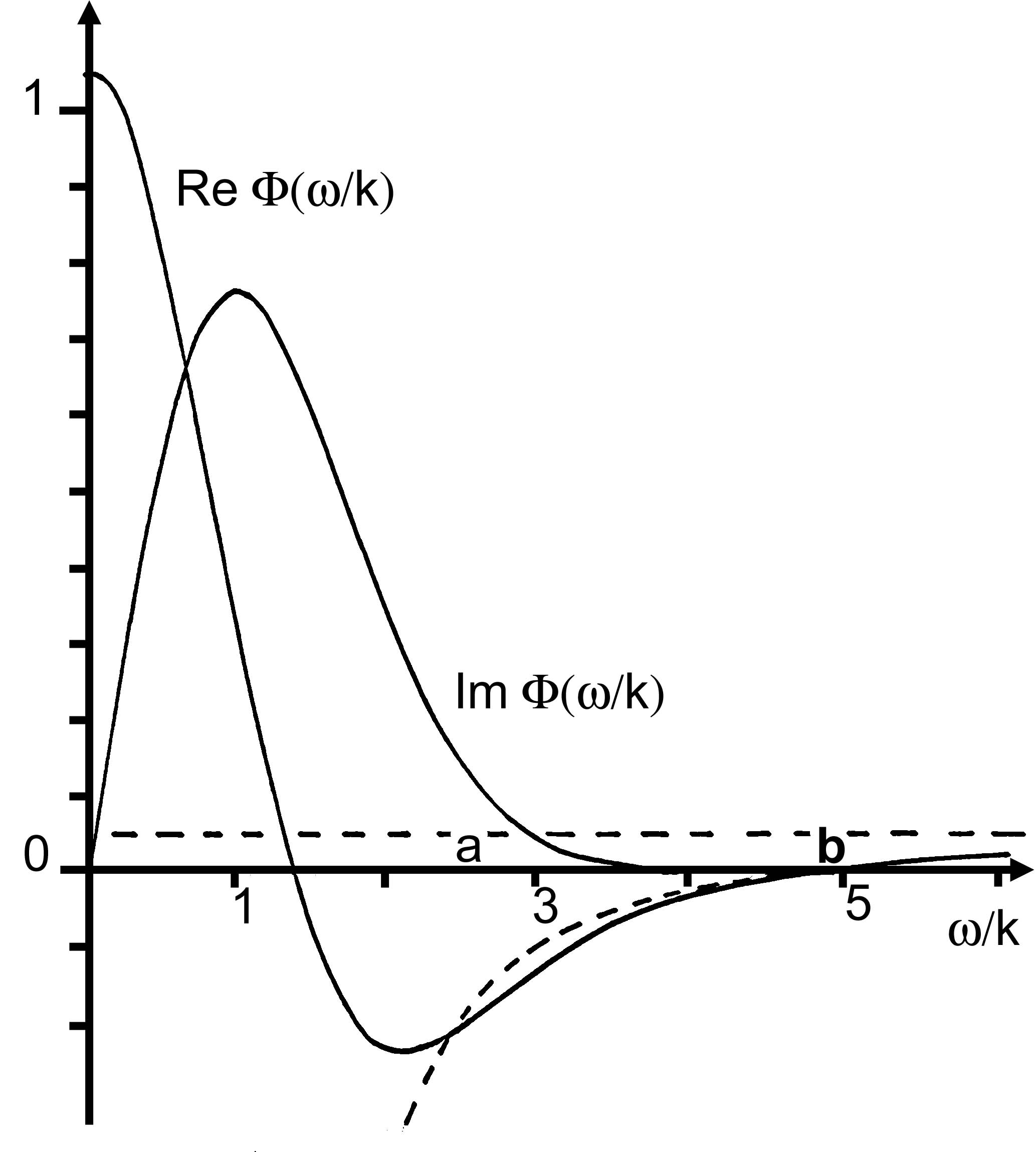,width=2.5in,angle=0}}
	\vspace*{1pt}
	\caption{The polarization function $\Phi(\omega/k)$ (in units such 
	that $\ompi=\kDi=1$) for a Maxwellian ion distribution and electron 
	Debye constant $k_{\rm De}^2 = 0.5$ (dashed horizontal line). The asymptotic expansion
	$\kDe^2 - (k^2/\omega^2 + \overline{v^2_{\rm i}}k^4/\omega^4)$ is 
	included for comparison, being represented by the dashed curve.
	\label{fig:MScFig2_2}}
\end{figure}

Although it is clearly impossible analytically to evaluate the
integral \eq{eq:screenedpotl} for general $g(v)$, or even for as
simple a distribution as the Lorentzian, one may derive
approximations for the integral in various ranges of $v_0$ that enable
one to gain some understanding of its properties.

First we sketch $\Phi(\omega/k)$ and its asymptotic expansion
in \fig{fig:MScFig2_2}.

\subsection{Small $v_0$}
\label{sec:3a_smallv0}

We note from equation \eq{eq:screenedpotl} that the argument of $\Phi
\equiv \Phi_{\rm e} + \Phi_{\rm i}$ is less than or equal to $v_0$.
Consquently for small $v_0$ only the behaviour of $\Phi$ near the
origin will affect the potential.

Expanding about the origin, $\Phi(w) = k_{\rm D}^2 + ia_0 w - a_1 w^2
+ O(w^3)$, where $a_0$ and $a_1$ are positive constants depending on
the distrubution function, we get, in units such that $k_{\rm D}^2
\equiv \kDi^2 + k_{\rm De}^2 = 1$,

\begin{eqnarray}
   \frac{1}{k^2+\Phi(w)}
   & = & \frac{1}{k^2+1} -\frac{ia_0 w}{(k^2+1)^2} \nonumber \\
    & &  \mbox{}+\left[\frac{a_1}{(k^2+1)^2} - 
    \frac{a_0^2}{(k^2+1)^3}\right]w^2 + \ldots
   \;.
   \label{eq:lovPhi}
\end{eqnarray}

Then
\begin{eqnarray}
    \varphi(\xvec)
    & = & \frac{q}{\varepsilon_0}\int\frac{d^3 k}{(2\pi)^3}
	e^{i\kvec\dotv\xvec}
	\left\{\frac{1}{k^2+1} 
	    - ia_0\frac{\khat\dotv\vvec_0}{(k^2+1)^2}
	\right.\nonumber  \\
	& & \phantom{q\int\frac{d^3 k}{(2\pi)^3} e^{i\kvec\dotv\xvec}}
	\left.
	    + \left[\frac{a_1}{(k^2+1)^2}
	    - \frac{a_0^2}{(k^2+1)^3}\right]
		  \vvec_0\dotv\khat\khat\dotv\vvec_0
	\right\}
	\nonumber  \\
    & = &
	\frac{q}{\varepsilon_0}\left\{\frac{\exp(-r)}{4\pi r}
	- ia_0\vvec_0\dotv\xhat\int\frac{d^3 k}{(2\pi)^3}
		  \frac{\xhat\dotv\khat }{(k^2+1)^2}e^{i\kvec\dotv\xvec}
	  \right.
	\nonumber  \\
    &  & \left.\mbox{}
	 +\vvec_0\vvec_0:
	 \left[
	 \int\frac{d^3 k}{(2\pi)^3}\khat\khat\; e^{i\kvec\dotv\xvec}
	     \left(
	     	\frac{a_1}{(k^2+1)^2}
	      - \frac{a_0^2}{(k^2+1)^3}
	     \right)
	 \right]
     \right\}
    \label{eq:varphi1}
\end{eqnarray}
where $\hat{\dotv}$ denotes a unit vector.

The first integral in the second line of \eq{eq:varphi1} may be 
evaluated as follows
\begin{eqnarray}
    \int\frac{d^3 k}{(2\pi)^3}
    \frac{\xhat\dotv\khat }{(k^2+1)^2}e^{i\kvec\dotv\xvec}
     & = & 
     \frac{1}{(2\pi)^2}\int_{-1}^{1}d\mu
     \int_{0}^{\infty}dk\frac{k^2}{(k^2+1)^2}e^{ik\mu r} \nonumber  \\
     & = & \frac{1}{i(2\pi)^2}\Re 
     \frac{d}{dr}\left(\frac{1}{r}-\frac{d}{dr}\right)\eta(r) \; ,
     \label{eq:integral1}
\end{eqnarray}
where $\eta(r)$ is defined as follows
\begin{equation}
    \eta(z) \equiv \frac{z}{i}\int_{0}^{\infty}dx\,\frac{\exp(ix)}{x^2+z^2}
    \label{eq:etadef}
\end{equation}
for $|\arg z| < \pi/2$ and by analytic continuation elsewhere, cutting 
the complex plane along the negative imaginary axis. For details of the properties of $\eta$ see Appendix I of the thesis.\cite{dewar67}

The tensor term in  the second line of \eq{eq:varphi1} may be 
evaluated by contour integration, yielding
\begin{eqnarray}
    \varphi(\xvec)
    & = & \frac{q}{\varepsilon_0}
        \left\{\frac{\exp(-r)}{4\pi r} 
	    - a_0\frac{\xhat\dotv\vvec_0}{(2\pi)^2}
	      \Re\frac{d}{dr}\left(\frac{1}{r}-\frac{d}{dr}\right)\eta(r)
	\right.\nonumber  \\
	& & \left.\phantom{\frac{1}{1}}\mbox{}
	 +\vvec_0\dotv
	 \left[
	     {\cal T}_{\parallel}\xhat\xhat
	    +{\cal T}_{\perp}(\Id - \xhat\xhat)
	 \right]\dotv\vvec_0
     \right\} \;,
    \label{eq:varphi2}
\end{eqnarray}
where 
\begin{eqnarray}
    {\cal T}_{\parallel}
    & \equiv & \frac{1}{8\pi}
    \left\{a_1
    \left[
	\left(
	    1+\frac{2}{r}+\frac{4}{r^2}+\frac{4}{r^3}
	\right)e^{-r}
	-\frac{4}{r^3}
    \right]
    \right.
    \nonumber \\
    & & \quad
    \left.\mbox{}
	- \frac{a_0^2}{4}
	\left[
	    \left(
		r+3+\frac{8}{r}+\frac{16}{r^2}+\frac{16}{r^3}
	    \right)e^{-r}
	    -\frac{16}{r^3}
	\right]
     \right\} \;,
    \label{eq:Tpardef} \\
    {\cal T}_{\perp}
    & \equiv & \frac{1}{8\pi}
    \left\{a_1
	\left[
	    \frac{2}{r^3}
	  -\frac{1}{r}
	    \left(
		1+\frac{2}{r}+\frac{2}{r^2}
	    \right)e^{-r}
	\right]
    \right.
    \nonumber \\
    & & \quad
    \left.\mbox{}
	- \frac{a_0^2}{4}
	\left[
	\frac{8}{r^3}
	   -\left(
		1+\frac{4}{r}+\frac{8}{r^2}+\frac{8}{r^3}
	    \right)e^{-r}
	\right]
     \right\} \;.
    \label{eq:Tperpdef}
\end{eqnarray}
It is to be noted that, contrary to appearances, ${\cal T}_{\parallel}$ 
and ${\cal T}_{\perp}$ are regular at $r = 0$.

For large $r$, the asymptotic form of $\eta(r) \sim 1/r$ holds and the 
exponential terms may be neglected. Thus, for $r \rightarrow \infty$,
\begin{eqnarray}
    \varphi(\xvec) & \sim & \frac{q}{\varepsilon_0}
    \left\{
	\frac{a_0}{\pi^2}\frac{\xhat\dotv\vvec_0}{r^3}
	+\frac{(a_0^2-a_1)}{2\pi}\frac{(\xhat\dotv\vvec_0)^2}{r^3}
    \right.\nonumber  \\  &  & 
    \left.\phantom{\frac{a_0}{\pi^2}\frac{\xhat\dotv\vvec_0}{r^3}}
	-\frac{(a_0^2-a_1)v_0^2}{4\pi}
	\frac{[1-(\xhat\dotv\hat{\vvec}_0)^2]}{r^3}
    \right\}
    \; .
    \label{eq:varphi3}
\end{eqnarray}

The first term is dominant for $|\xhat\dotv\vvec_0| \gg 0$ i.e. for
all directions not oblique to the direction of motion.  The field is
reminiscent of a dipole field except that it decays more rapidly with
$r$.  This may qualitatively be interpreted as meaning that the centre
of the screening cloud has been displaced to a position behind the
particle.  It is this asymmetry that gives rise to the drag on a very
heavy particle moving at subthermal speeds and, since this drag is the
summation of the field of each individual particle moving in the
average field of all the others, one supposes that the drag thus
calculated includes single particle effects to the extent of validity
of the linearized theory,\footnote{This is confirmed by the work of
Hubbard\cite{Hubbard_1961a} who shows that the friction coefficient is the
sum of the ``self field" term and a fluctuating microfield term which
vanishes in the limit of infinite test-particle mass, as do the higher
transition moments.} i.e. except for very close collisions.

Rand\cite{rand59} has calculated the screened field to first order in
$v_0$ by considering the trajectories of individual particles,
explicitly rather than through the Vlasov equation.  He obtains a
function $\phi_1(x)$ as a triple integral, which he apparently
evaluates numerically but which agrees exactly\footnote{There is an
error in Eq.  (26) of Ref.  2, namely that the second $Z$ in the
equation should be replaced by $2Z/(Z+1)$.} with values calculated
from $\frac{1}{2}{\rm Re}(d/dx)(d/dx - 1/x)\eta(x)$.

It is interesting to note that the solution we have obtained differs
completely from that of Pines and Bohm\cite{pines-bohm52} and workers
who have used linearized fluid dynamical
equations.\cite{majumdar60,cohen61}  We ascribe this discrepancy to
these authors' use of a collective approach in a region in which it is
not valid:
Inspection of \fig{fig:MScFig2_2} shows
that the region in which collective coordinates exist can only 
be $|\omega/k| \gtrsim \bar{v}_{\rm i}^2$, which
is never entered in the case of a slow particle. The criterion
$k \lesssim \kDi$ is not adequate in the case of forced oscillations since
$\omega$ may be much less than $\ompi$. 

We note that behind the test particle the potential has opposite sign
to its unscreened value, which we refer to as positive.  At right
angles to the direction of motion the first order term vanishes and
the second-order term dominates.  We see that $\varphi(\xvec)$ goes
negative in this direction if $a_0^2 > a_1$.  For the Lorentzian
distribution we have from \eq{eq:Phi_Lor} $\Phi_i(x) =
1+2ix-3x^2+\kDe^2+O(x^3)$, in units where $\ompi = \kDi =
1$.  Thus $a_0 = 2$ and $a_1 = 3$, and therefore $a_0^2 > a_1$ so that
the $90^{\circ}$ potential does go negative somewhere between $r=0$
and $r= \infty$.

We shall now discuss the situation when the particle has 
sufficient velocity for wave excitation to be possible.

The dispersion relation for plasma waves is $\epsilon(\omega,k) = 0$, i.e.
\begin{equation}		
    \Phi(\omega/k) = -k^2 \;,
    \label{eq:disprel}
\end{equation}
where $\omega$ is in general complex, but if the waves are but slightly 
damped then we may take $\omega$ real as a first approximation. It is then
clear that only the regions in which $\Re\Phi(\omega/k)$ is negative are 
available for excitation. In \fig{fig:MScFig2_2} this means the region
$a < \omega/k < b$ which, it will be noted, exists if
$\kDe^2$ is sufficiently small.

This is the region of ion acoustic waves, which resemble sound waves in ordinary gases in that 
there is an upper bound, $b$, to their phase velocity. To guide our 
mathematical investigations we shall endeavour to give a physical 
picture of of the processes that can occur.

\begin{figure}[th]		
	\centerline{\psfig{file=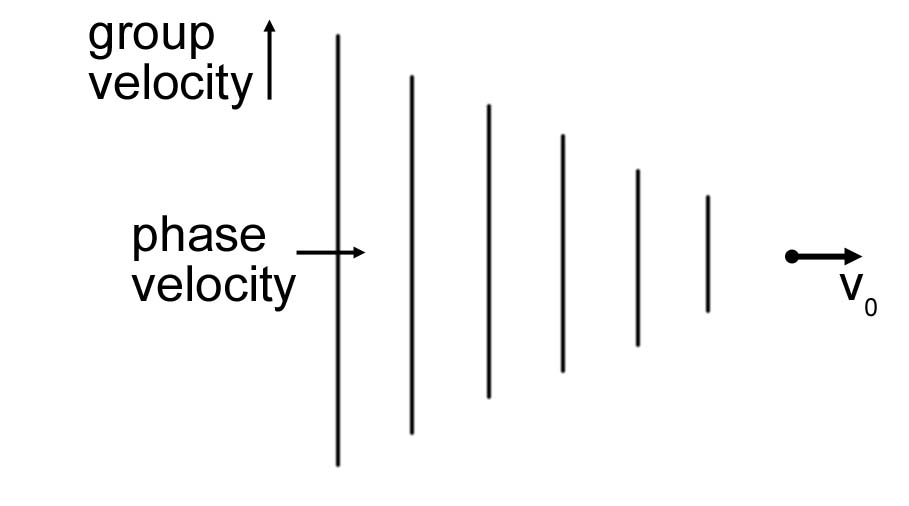,width=2.5in,angle=0}}
	\vspace*{1pt}
	\caption{Schematic of wake field behind a charged test particle 
	moving through a plasma.
	\label{fig:MScFig2_3}}
\end{figure}

If a particle has a velocity $v_0$, $a < v_0 < b$, then it can
excite waves with the same phase velocity as its own velocity. 
We might therefore expect it to be followed by a train of waves with
this phase velocity, spreading out laterally with the maximum group
velocity possible. This is roughly indicated in \fig{fig:MScFig2_3}.

If $v > b$ then we can expect no monochromatic train but some kind of
shock wave confined within the ``Mach cone''.  This is analogous to
the formation of the shock wave behind a supersonic object in
the atmosphere or to a longitudinal Cerenkov radiation.

A note of caution must be sounded at this point with respect to
the spreading out of the wake.  Since this is controlled by
the interference of the generated waves it is critically
dependent on the dispersive nature of the medium, as will be
discussed later. It
is well known that in regions of anomalous dispersion in optical media
the concept of group velocity breaks down entirely.\cite[p. 333]{stratton41} We thus
expect the situation to be complicated if Landau damping becomes important.

Since we are seeking wave behaviour a contour integral method
is the obvious choice and has been	much used, see e.g. Ref.~\refcite{pines-bohm52}. We
indicate the general method below as well as some of its analytical difficulties, not mentioned in the literature due in effect to the assumption from the outset that an
asymptotic form for the dielectric constant is a valid approximation.

In equation \eq{eq:screenedpotl} use cylindrical coordinates with axis parallel to $\vvec_0$. Then
\begin{eqnarray}		
    \varphi & = &
    \frac{q}{(2\pi)^2\varepsilon_0}
   \int_{0}^{\infty}\!\!\!\! dk_{\perp}\,k_{\perp}J_{0}(k_{\perp}x_{\perp}) \times \nonumber \\
   & & 
   \int_{-\infty}^{\infty}\!\!\!\!dk_{1} \exp (i k_1 x_1)
                                                         \left[k_1^2 + k_{\perp}^2
                                                         + \Phi\left(\frac{k_1 v_0}{\left(k_1^2 + k_{\perp}^2 \right)^{1/2}}\right)
                                                         \right]^{-1}
   \;.
    \label{eq:screenedpotlCyl}
\end{eqnarray} 

\begin{figure}[th]		
	\centerline{\psfig{file=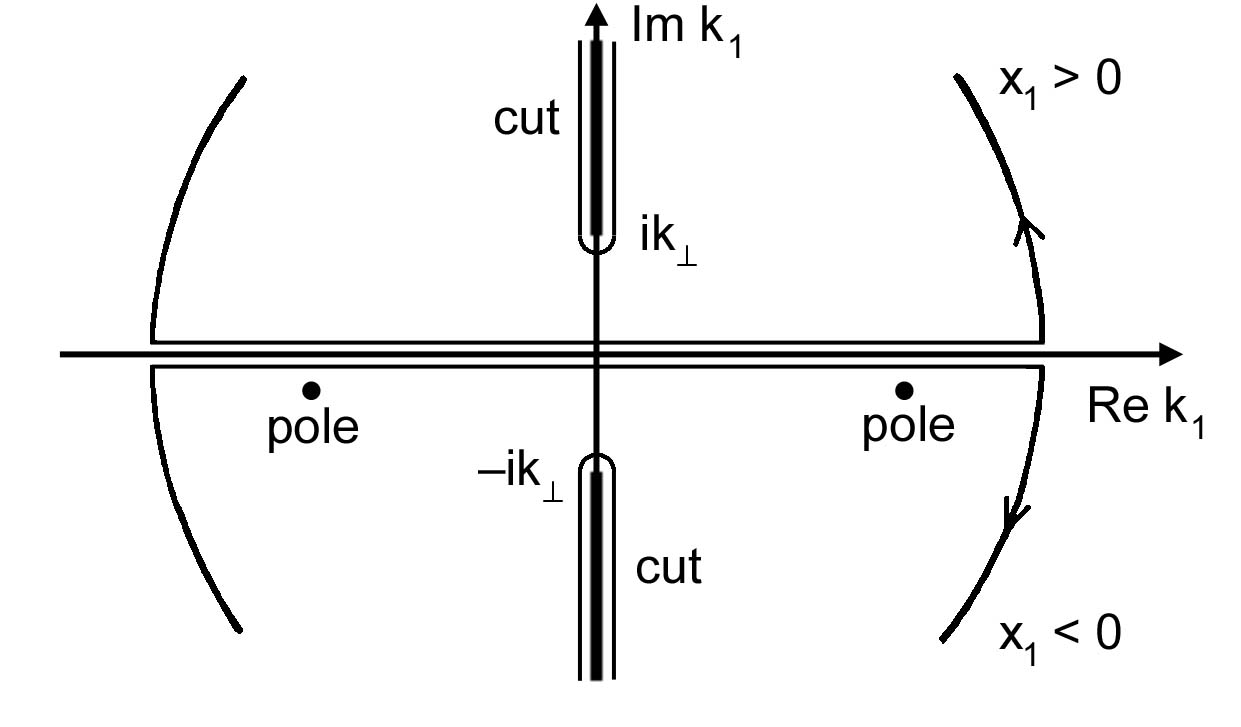,width=2.5in,angle=0}}
	\vspace*{1pt}
	\caption{Contours used in the evaluation of the $k_1$ integral in \eq{eq:screenedpotlCyl}.
	\label{fig:MScFig2_4}}
\end{figure}

The contour in the $k_{1}$ integral may be completed in the lower/upper half of the complex $k_{1}$  plane according as $x_{1} \lessgtr 0$. However, note that, since $k_{1} = \pm k_{\perp}$ are branch points, the contours must be indented as shown in \fig{fig:MScFig2_4}. The zeros of the denominator will give rise to poles whose contributions may be evaluated by the method of residues.

For reasonably stable distributions, at least, one may show that the plasma wave poles corresponding to \eq{eq:disprel} are in the lower half $k_1$ plane. Consequently there is a wave excitation only \emph{behind} the particle. There are other zeros of $\epsilon$ somewhere in the lower half plane, but since they give rise to strongly damped contributions, one hopes that they may be ignored, at any rate for large $|x_1|$. Indeed the contribution from the region of the cut will also be negligible compared with the wave contribution.

For $x_1 > 0$ it is necessary to consider the contribution from the region of the cut. The part nearest the real axis is clearly the most important for large  $|x_1|$, so let us consider the region around $ik_{\perp}$

In this region $|k_1v_0/(k_1^2 + k_{\perp}^2)^{1/2}|$ is large and the asymptotic expansion for $\Phi$ may be used. This may be obtained from \eq{eq:Phiapprox} by expanding the denominator and integrating term by term.

\begin{equation}		
    \Phi(z) = 
    \kDe^2 - \ompi^2\left(\frac{1}{z^2} + \frac{\overline{v_{\rm i}^2}}{z^4} \right)
    -2\pi i\theta(-\Im z)g'_{\rm i}(z)\ompi^2
   \;,
    \label{eq:PhiAsymp}
\end{equation} 
where the last term is required for analytical continuation into the lower half $z$ plane. Here $\overline{v_{\rm i}^2}$ is the mean-square ion speed (the square of the \emph{ion thermal speed} in a Maxwellian plasma).
However, if $k_1$ is in the lower/upper half plane, so also is $k_1/\left(k_1^2 + k_{\perp}^2 \right)^{1/2}$.
Hence the analytical continuation term is not required in the region we are considering. At $k_1 = -ik_{\perp}$ it gives rise in general to an essential singularity, but this contribution is being neglected as
mentioned above. We now note the curious fact that the asymptotic expansion is single-valued in the upper half plane since only even
powers of $\left(k_1^2 + k_{\perp}^2 \right)^{1/2}$  occur. Thus there is only a simple pole in the neighbourhood of $k_1 = ik_{\perp}$.

For simplicity let us retain only terms to order $1/z^2$. Note that to this order the maximum phase velocity is
\begin{equation}		
    b = \ompi/\kDe \equiv C_{\rm s}
   \;,
    \label{eq:vpmax}
\end{equation} 
the constant $C_{\rm s}$ being commonly called the \emph{ion sound speed}.

\subsection{Intermediate velocity $(\overline{v_{\rm i}^2})^{1/2} \ll v_0 < C_{\rm s}$ (forward field)}
\label{sec:3bi_intermediatev0}

If $v_0 \ll b$ then the pole is approximately at $k_1 = ik_{\perp}$ and we may take $\kDe = 0$ without greatly altering the situation.
Then
\begin{equation}		
    \varphi =  \frac{q}{4\pi\varepsilon_0}\left[
    \frac{1}{r}
    - \frac{\ompi}{v_0}\int_0^{\infty}\!\!\!dy\,
    \frac {J_0\left(\ompi x_{\perp}y/v_0\right)\exp\left(-\ompi x_1 y/v_0\right)}{1 + y^2}
    \right]
   \;.
    \label{eq:screenedpotlInter}
\end{equation} 

An alternative form is obtained by noting that\cite{watson44}
\begin{eqnarray}		
    \frac{1}{r} & = & \int_0^{\infty} \!\! dk_{\perp}\, J_0(k_{\perp}x_{\perp}) \exp(-x_1 k_{\perp})  \nonumber\\
    & = & \frac{\ompi}{v_0}\int_0^{\infty} \!\!  dy\, J_0\left(\frac{\ompi x_{\perp}y}{v_0}\right)
    \exp\left(-\frac{\ompi x_1 y}{v_0}\right)
    \;.
    \label{eq:WatsonIdent}
\end{eqnarray} 
Thus
\begin{equation}		
    \varphi =  \frac{q}{4\pi\varepsilon_0}\frac{\ompi}{v_0}
    \int_0^{\infty}\!\! dy\,
    \frac {y^2}{1 + y^2}J_0\left(\frac{\ompi x_{\perp}y}{v_0}\right)\exp\left(-\frac{\ompi x_1 y}{v_0}\right)
   \;.
    \label{eq:screenedpotlAlt}
\end{equation} 

Note that the characteristic length is now $v_0/\ompi$, the distance the particle moves in an ion plasma oscillation time, rather than the ion Debye length. The new length is longer, and furthermore, the screening is again not exponential at large $r$ as we shall now show.

\begin{eqnarray}		
\mbox{For} \: x_{\perp}  =  0: \:   \varphi & = &  \frac{q}{4\pi\varepsilon_0}\frac{1}{x_1}\left[
    1 - \frac{\ompi x_1}{v_0} i \eta\left(\frac{i\ompi x_1}{v_0}\right)
    \right] \nonumber \\
    & \sim &  \frac{q}{4\pi\varepsilon_0}\frac{2\ompi}{v_0}\left(\frac{\ompi r}{v_0}\right)^{-3}
   \;.
    \label{eq:screenedpotlxp0} \\
\mbox{For} \: x_1 = 0: \:   \varphi & = &  \frac{q}{4\pi\varepsilon_0}\frac{1}{x_{\perp} }\left\{
    1 - \frac{\ompi x_{\perp} }{v_0}\left[
    I_0\left(\frac{\ompi x_{\perp}}{v_0}\right) - {\rm\bf L}_0\left(\frac{\ompi x_{\perp}}{v_0}\right) 
    	\right]
    \right\}\nonumber \\
    & \sim & - \frac{q}{4\pi\varepsilon_0}\frac{\ompi}{v_0} \left(\frac{\ompi r}{v_0}\right)^{-3}
   \;,
    \label{eq:screenedpotlx10}
\end{eqnarray} 
where $I_0$ and ${\bf L}_0$ modified Bessel and Struve functions respectively.

A general asymptotic form at large $r$ may be obtained from \eq{eq:screenedpotlAlt}  by approximating the denominator to 1 and differentiating \eq{eq:WatsonIdent} twice with respect to $\ompi x_1/v_0$. Thus
\begin{equation}	
	\varphi \sim  \frac{q}{4\pi\varepsilon_0}\frac{\ompi}{v_0}\frac{2(\hat{\xvec}\dotv\hat{\vvec}_0)^2
	- (\hat{\xvec}\cross\hat{\vvec}_0)^2}{(\ompi r/v_0)^3}
\label{eq:screenedpotlAsymp}
\end{equation}

Note that this obeys the same inverse third power law as in the very low velocity case, and also as in this case the field goes negative, for large $r$, at large angles between $\hat{\xvec}$ and $\hat{\vvec}_0$. The angle at which the field changes sign is approximately $55^{\circ}$.

\subsection{Supersonic velocities $v_0 > C_{\rm s}$ (forward field)}
\label{sec:3ci_supersonicv0}

As $v_0$ increases beyond maximum phase velocity $b$ (the ion sound speed) the pole moves from $ik_{\perp}$ to $i\kDe$ and the forward field changes to a Debye potential with Debye length $\kDe^{-1}$. This has the physical interpretation that if the particle velocity greatly exceeds the maximum ion wave phase velocity then the ions are too sluggish to participate in the screening.

We must now discuss the case $x_1 < 0$ where plasma wave excitation is assumed to be the dominant contribution. The positions of the plasma wave poles are given by the dispersion relation \eq{eq:disprel}.
It will be assumed that the asymptotic expansion \eq{eq:PhiAsymp} represents a valid approximation to $\Phi$, with the one alteration that, since $z$ is on or near the real axis, the last term is $-i\pi g'_{\rm i}(z)$.
[It will be noted that in a sector containing the real axis $g'_{\rm i}(z)$ usually decays exponentially, and the factor multiplying it is irrelevant to the \emph{asymptotic expansion};
but we require an \emph{approximation} to $\Phi$ and so analytically continue off the real axis where the $i\pi$ factor is known to be exact.]

The dispersion relation is now
\begin{equation}	
	k^2 + \kDe^2 - \ompi^2 \left(\frac{k^2}{\omega^2} + \overline{v_{\rm i}^2} \frac {k^4}{\omega^4}\right)
	- i\pi \ompi^2 g'_{\rm i}\left(\frac{\omega}{k}\right) = 0 \;.
\label{eq:disprelapprox1}
\end{equation}

This is valid for $|\omega/k| \gg (\overline{v_{\rm i}^2})^{1/2}$, i.e. when the ``finite temperature'' term $v_{\rm i}^2k^4/\omega^4$ is small compared with $k^2/\omega^2$.

\subsection{Intermediate velocity $(\overline{v_{\rm i}^2})^{1/2} < v_0 < C_{\rm s}$ (wake)}
\label{sec:3bii_intermediatev0}

We again make the simplifying assumption that $\kDe = 0$. The dispersion relation can now be reduced to the familiar form
\begin{equation}	
	\omega^2 \approx \ompi^2 + \overline{v_{\rm i}^2} k^2
	+ i\pi\ompi^2\sgn\omega\left(\frac{\ompi}{k}\right)^2 g'_{\rm i}\left(\frac{\ompi}{k}\right)
	= 0 \;.
\label{eq:disprelapprox2}
\end{equation}

To simplify the analysis let us use units such that $\kDi = \ompi = 1$. In these units $ \overline{v_{\rm i}^2}  = 3$ (Maxwellian case), and generally we may assume $ \overline{v_{\rm i}^2}  \sim 1$.

The finite temperature correction is small provided $k^2 \ll 1$, i.e. for wavelengths much greater than a Debye length. Thus, for the dispersion relation to be valid, we must require both $|k_1| \ll 1$
and $k_{\perp} \ll 1$. Since $\omega = k_1 v_0 \approx 1$ it is clear that we require $v_0 \gg 1$.

Substituting $\omega = k_1 v_0$ and $k^2 = k_1^2+ k_{\perp}^2$, and assuming $k_{\perp}$ small, we find
\begin{eqnarray}	
	k_1 & = &\pm\frac{1}{v_0}
	\left\{ 
		1 + \frac{\alpha}{2} + \alpha v_0^2 k_{\perp}^2
		\phantom{\left[\frac{1+\alpha}{v_0^2} + k_{\perp}^2\right]^{-1}}\right. \nonumber\\
	&& \phantom{\pm\frac{1}{v_0}}
	\left.	\pm \frac{i\pi}{2}\left[\frac{1+\alpha}{v_0^2} + k_{\perp}^2\right]^{-1}
		g'\left(\left[\frac{1+\alpha}{v_0^2} + k_{\perp}^2\right]^{-1/2}\right)
	\right\} \;,
	\label{eq:k1approx}
\end{eqnarray}
where $\alpha \equiv v_{\rm i}^2/v_0^2 \ll 1$.

The last term takes account of Landau damping and, $g'(v)$ being negative, serves to displace the poles in \fig{fig:MScFig2_4} below the real $k_1$ axis.
Neglecting the damping term, which is very small until $k_{\perp}$ gets close to 1, we find
\begin{eqnarray}	
	&&\frac{\partial}{\partial k_1}\left[
		k_1^2 + k_{\perp}^2 + \Phi\left(\frac{k_1 v}{\left(k_{\perp}^2 + k_1^2\right)^{1/2}}\right)
	\right]_{\rm poles} \nonumber\\
	&& \phantom{\frac{\partial}{\partial k_1}(k_1^2 + k_{\perp}^2)}\quad\quad \quad
	= 
	\pm\frac{2}{v}\left(1 + \frac{\alpha}{2}\right)\left(1 + \frac{\alpha}{2}y^2\right)\left(1 + y^2\right) \;,
	\label{eq:poles}
\end{eqnarray}
where $y \equiv k_{\perp}v_0$.

The two factors containing $\alpha$ are small correction terms. Using \eq{eq:k1approx} to locate the poles and \eq{eq:poles} to evaluate the residues we find from \eq{eq:screenedpotlCyl} that
\begin{equation}	
	\varphi  =  \frac{2-\alpha}{v_0}\,\Im\left[
	\exp\left(-i\left[1+\frac{\alpha}{2}\right]\left|\frac{x_1}{v_0}\right|\right)
	I\left(\frac{\alpha}{2}\left|\frac{x_1}{v_0}\right|,\frac{x_{\perp}}{v_0},\frac{\alpha}{2}\right)
	\right] \;,
	\label{eq:residues}
\end{equation}
where
\begin{equation} 
	\gamma(y) \equiv -\frac{\pi}{2}\frac{v_0^2}{(1+\alpha+y^2)}\, g'\left(\frac{v_0}{(1+\alpha+y^2)^{1/2}}\right)
	\label{eq;gammadef}
\end{equation}
is the Landau damping term, which has the effect of rapidly cutting off the integral when $|y|$ gets close to $v_0$, and $I$ is the integral
\begin{equation} 
	I(a,b,c) \equiv \int_0^{\infty}\!\!\!dy\,y\,
	\frac{J_0(by)\exp(-iay^2)}{(1+y^2)(1+cy^2)}\psi(y) \;,
	\label{eq:integraldef}
\end{equation}
with
\begin{equation} 
	\psi(y) \equiv \exp\left(-\left|\frac{x_1}{v_0}\right|\gamma(y)\right) \;.
	\label{eq:smallpsidef}
\end{equation}
In \eq {eq:residues} the subsitutions for the dummy arguments $a$, $b$ and $c$ are
\begin{equation}
	a = \frac{\alpha}{2}\left|\frac{x_1}{v_0}\right|\;, \quad
	b = \frac{x_{\perp}}{v_0} \quad
	\mathrm{and}\quad c = \frac{\alpha}{2} \;.
	\label{eq:abcvals}
\end{equation}

When $b=0$ we suppose that the cutoff factor $\psi$ may be neglected. Then
\begin{equation} 
	I(a,0,c) \approx \frac{1}{2(1-c)}\left[e^{ia}E_1(ia) - e^{ia/c}E_1\left(\frac{ia}{c}\right)\right]
	\label{eq:Ib0}
\end{equation}
where $E_1$ is the exponential integral.\cite{Abramowitz_Stegun_72}

To obtain an asymptotic form in the general case it is helpful to transform the one-sided integral into a two-sided integral by a method due to Hankel,\cite{watson44} noting that, for $x>0$,
\begin{equation}
	J_0(x) = \frac{1}{2}\left[H^{(1)}_0(x) - H^{(1)}_0(-x + i0) \right] \;,
	\label{eq:J0ident}
\end{equation}
where $H^{(1)}_0(z)$ is a Hankel function, defined on the complex $z$-plane cut along the negative real axis, being regular elsewhere. Then
\begin{equation} 
	I(a,b,c) = \frac{1}{2}\int_{-\infty}^{\infty}\!\!\!dy\,y\,
	\frac{H^{(1)}_0(b[y+i0])\exp(-iay^2)}{(1+y^2)(1+cy^2)}\psi(y) \;.
	\label{eq:integral2side}
\end{equation}

If $a$ were zero and $\psi$ well behaved, we could evaluate the integral by completing the contour in the upper half plane. Instead we seek the saddle points and poles of the integrand, the poles being at $y=\pm i$ and $y = \pm i/\sqrt{c}$. To find the saddle points let us suppose that the asymptotic form for $H^{(1)}_0(z)$,
\begin{equation} 
	H^{(1)}_0(z) \sim \sqrt{\frac{2}{\pi}}\,\frac{\exp i(z - \frac{\pi}{4})}{\sqrt{z}}\;, \quad |z| \rightarrow 0 \;,
	\label{eq:Hankelasymp}
\end{equation}
is valid in the region of these points. (The asymptotic form will be valid in the region of the poles provided $|b| \gg 1$, $c<1$.) 

Since the denominator is slowly varying away from its zeros the saddle points are determined by the exponential factor, the exponent of which, being quadratic, has only one stationary point. This is at $y = y_0$, where $y_0 = b/2a$. For \eq{eq:integral2side} to be valid we require $|y_0| \gg 1$.\

We conclude that, provided the two conditions $b \gg 1$ and $b \gg 2|a|$ are satisfied, we may approximate $I$ by
\begin{equation} 
	I(a,b,c) \approx \frac{e^{-i\pi/4}}{\sqrt{b}}\sqrt{2\pi}\int_{-\infty}^{\infty}\!\frac{dy}{2\pi}\,
	\frac{(y+i0)^{1/2}\exp(-iay^2)\psi(y)}{(1+y^2)(1+cy^2)}e^{iby} \;.
	\label{eq:integralapprox}
\end{equation}

This is a Fourier transform, which may be transformed by the convolution theorem to
\begin{equation} 
	I(a,b,c) \approx \frac{e^{-i\pi/4}}{\sqrt{b}}\frac{1}{\sqrt{2\pi}}\,\psi_b*\int_{-\infty}^{\infty}\!\!\!dy\,
	\frac{(y+i0)^{1/2}\exp(-iay^2)e^{iby}}{(1+y^2)(1+cy^2)} \;,
	\label{eq:integralconv}
\end{equation}
where $\psi_b \equiv \int_{-\infty}^{\infty}\psi(y)e^{iby}dy/2\pi$ is a delta-like function of $b$ with width $1/v_0$, which will have the effect of destroying all wave structure with wavelength of a Debye length or less.

The integral may now be estimated by deforming the contour of integration to cross the saddle point along the line of steepest descent. (The pole at $i/\sqrt{c}$, being far up the imaginary axis, gives negligible contribution.)
\begin{figure}[h!]		
	\centerline{\psfig{file=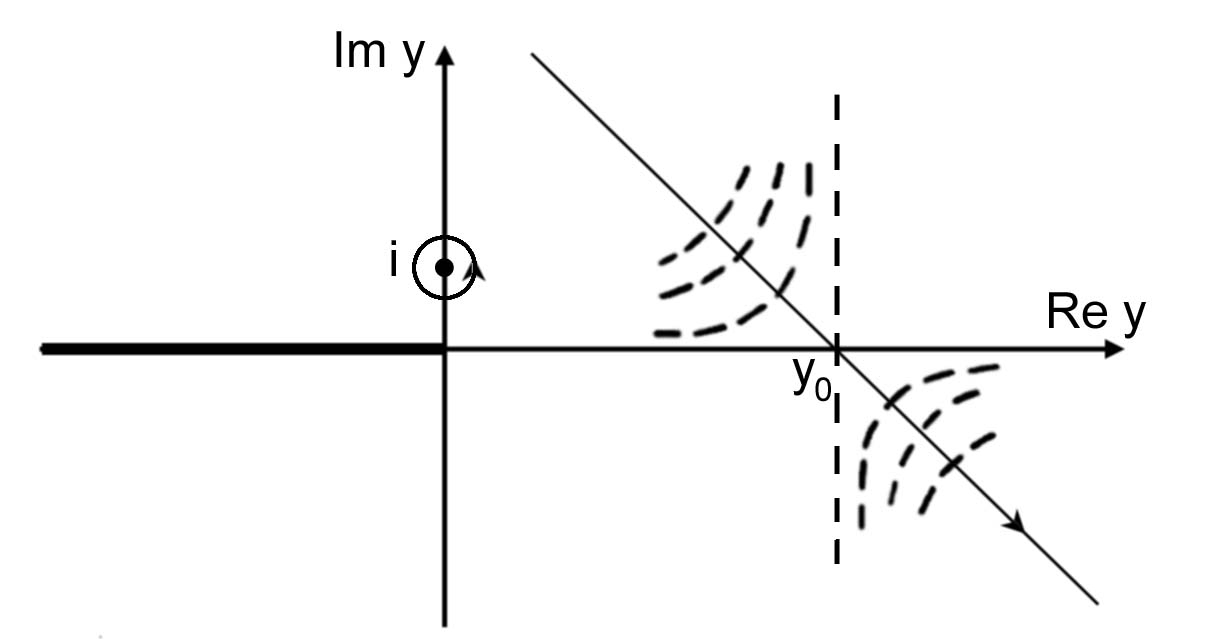,width=3in,angle=0}}
	\vspace*{1pt}
	\caption{The contour used for estimating the integral in \eq{eq:integralconv}.
	\label{fig:MScFig2_5}}
\end{figure}

We make the further requirement that the width of the saddle at $y_0$ be much less than $y_0$. That is, $1/\sqrt{a} \ll y_0$. Then
\begin{equation} 
	I(a,b,c) \approx  \psi_b* \left[
	\sqrt{\frac{\pi}{2b}}\,\frac{e^{ia-b}}{1-c}
	-\frac{i}{\sqrt{2ab}}\frac{y_0^{1/2}\exp(ib^2/4a)}{(1+y_0^2)(1+cy_0^2)} \right] \;.
	\label{eq:integralsaddle1}
\end{equation}

Since the first term varies slowly compared with $\psi_b$, the convolution will leave it but little changed, whereas the rapid fluctuations in $\exp(ib^2/4a)$ will be damped out. Supposing that the exponential term may locally be approximated by a monochromatic wave, we finally obtain
\begin{equation} 
	I(a,b,c) \approx  
	\sqrt{\frac{\pi}{2b}}\,\frac{e^{ia-b}}{1-c}
	-\frac{i}{\sqrt{2ab}}\frac{y_0^{1/2}\exp(ib^2/4a)}{(1+y_0^2)(1+cy_0^2)} \;,
	\label{eq:integralsaddle2}
\end{equation}
with $y_0 = b/2a$.

The method of steepest descent we have used to approximate the integral is equivalent, in the neighbourhood of the saddle point, to the method of stationary phase employed by Majumdar\cite{majumdar63} but takes better account of the behaviour away from the saddle point by including the contribution of the pole at $y = i$. We have also attempted to take some account of Landau damping, which the above author was unable to do owing to his formulation in terms of fluid dynamics.

Below we recapitulate the criteria for the validity of \eq{eq:integralsaddle2}
\begin{equation} 
	c \ll 1, \quad b \gg 1, \quad b \gg 2a, \quad b \gg 2\sqrt{a}\;.
	\label{eq:inequalities}
\end{equation}
With $a$, $b$ and $c$ given by \eq{eq:abcvals} these correspond physically (recalling that in this subsection we are using units such that $\kDi = \ompi = 1$) to the inequalities
\begin{equation}
	\frac{\alpha}{2} \ll 1, \quad \frac{x_{\perp}}{v_0} \gg 1, \quad x_{\perp} \gg \alpha|x_1|, \quad x_{\perp} \gg (\overline{v_i^2}|x_1|)^{1/2} \;.
	\label{eq:physinequalities}
\end{equation}

\begin{figure}[thb]		
	\centerline{\psfig{file=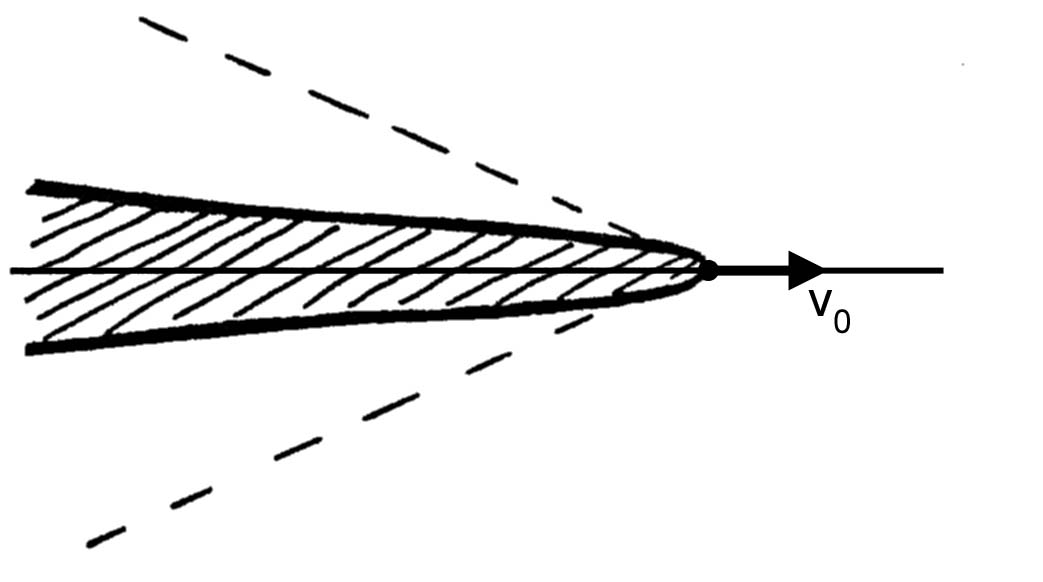,width=3in,angle=0}}
	\vspace*{1pt}
	\caption{Schematic of the wake region of an intermediate-velocity test particle indicating where the approximations used to derive \eq{eq:integralsaddle2} break down (hatched region) and the ``thermal Mach cone'' (dashed line).
	\label{fig:MScFig2_6}}
\end{figure}
Observe that the Landau damping term cuts off the second term at about $x_{\perp} = |x_1|/v_0$, which might be called a thermal Mach cone. This behaviour can be understood from the fact that plasma waves with high group velocity have low phase velocity, and hence are strongly damped. The first term, however, is unaffected by Landau damping and exhibits no thermal Mach cone, although it decays exponentially with $x_{\perp}$. We indicate the thermal Mach cone and the region excluded by the inequalities in \eq{eq:physinequalities} in \fig{fig:MScFig2_6}.

\subsection{Supersonic velocities $v_0 > C_{\rm s}$ (wake field)}
\label{sec:3cii_supersonicv0}

This is the case considered by Kraus and Watson\cite{kraus-watson58} but their treatment neglects finite ion temperature effects on the dispersion relation and is consequently inadequate for finding variations over distances less than $\kDe^{-1}$. We shall not go into detail on this case but shall indicate the approximations that may be made and the integral obtained.

The simplifying assumption made by the above authors can be represented as
\begin{equation}
	(\overline{v_i^2})^{1/2} \ll b \ll v_0 \;,
	\label{eq:SupersonicWakeineqs}
\end{equation}
where $b$ is here the upper bound to the phase velocity [$\approx C_{\rm s}$ by \eq{eq:vpmax}], not the dummy argument used in the previous subsection. Taking units such that $\ompi=\kDe=1$ we have
\begin{equation}
	\overline{v_i^2} \ll 1\;, \quad b \approx 1\;, \quad v_0 \gg 1 \quad \mathrm{and} \quad \kDi \gg 1 \;.
	\label {eq:SupersonicWakeineqsND}
\end{equation}

Using the same methods as before, but neglecting Landau damping, we find the approximate form for $x_1 < 0$
\begin{equation}
	\varphi \approx
	-\frac{2}{v_0}\int_0^{\infty}\!\!\!\frac{dk_{\perp}\,k_{\perp}^2}{(1 + k_{\perp}^2)^{3/2}}\,
		J_0(x_{\perp}k_{\perp})
		\sin\left(
				\left|\frac{x_1}{v_0}\right|\frac{k_{\perp}(1 + \frac{1}{2}\overline{v_i^2}k_{\perp}^2)}{(1 + k_{\perp}^2)^{1/2}}
			\right) \;.
	\label{eq:SupersonicWakeInt}
\end{equation}

Kraus and Watson in effect assume that the critical part of the integral for determining the large $|x_1|/v_0$ behaviour of ($x_{\perp}$ small) is near $k_{\perp} = 0$ so that terms such as $1 + k_{\perp}^2$ may be approximated to 1. However, it must be observed that there is a range of $k_{\perp}$ (between 1 and $1/\overline{v_i^2}$) in which the argument of the sine function is but slowly varying. This suggests that, superimposed on the endpoint contribution, there may be a sinusoidal term.

When $x_{\perp}$ is not small we may transform the integral into a two-sided integral as in Sec.~\ref{sec:3bii_intermediatev0}. There will be a saddle point which disappears into the origin at the Mach cone $x_{\perp} = |x_1|/v_0$, but the contribution from the singularities of the integrand will be complicated by the essential singularities at $k_{\perp} = \pm i$.

\section{Numerical Solution}
\label{sec:NumSol}

The explorations of the previous section showed that some insight into the nature of the fields around a particle in a collisionless plasma, as given formally by \eq{eq:response}, may be obtained by asymptotic analyses in various limits. However, these are often difficult and the errors difficult to quantify. It is thus essential to supplement such analytic work with numerical and graphical methods (and \emph{vice versa}).

In this section we summarize how this was done computationally in my MSc thesis work\cite{dewar67} using the newly arrived IBM 7044 (which had just replaced the Australian-built computer CSIRAC\cite{AlistairMoffat_06}).

\subsection{Formulation of the numerical method}
\label{sec:NumForm}

To evaluate the integral \eq{eq:response} numerically we must choose axes such that as much of the integration as possible may be done analytically while the remaining integrations are as tractable as possible. It is found that, although one may come tantalizingly close to reducing the triple integral to a single integral, it is in general necessary to evaluate a double integral.
\begin{figure}[thb]		
	\centerline{\psfig{file=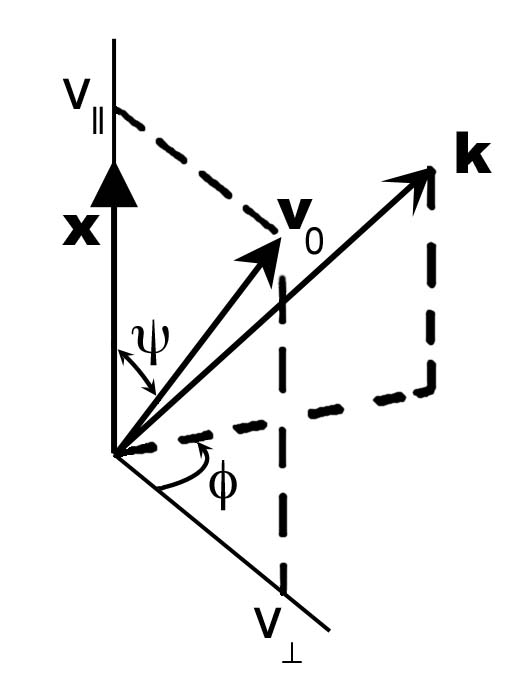,width=1.5in,angle=0}}
	\vspace*{1pt}
	\caption{Coordinate system used for numerical evaluation of the triple integral in \eq{eq:response}.
	\label{fig:MScFig3_1}}
\end{figure}

Three systems of coordinates suggest themselves: the cylindrical coordinates of Sec.~\ref{sec:3_Analapprox}, spherical polars with axis along $\vvec_0$, and spherical polars with axis along $\xvec$. The first method involves infinite integrals with integrands having a large peak near the plasma wave pole and we consequently reject it. The last choice appears to have the advantage over the remaining one that it is easier to understand the behaviour of the integrands (at least at low $\vvec_0$) and to isolate the singularity caused by the behaviour of $\eta(x)$ at $x = 0$. These axes are sketched in \fig{fig:MScFig3_1}. In this coordinate system we have
\begin{equation}
	\hat{\kvec}\dotv\vvec_0 = \mu v_{\parallel} + \sqrt{1-\mu^2}\cos\phi\,v_{\perp} \;,
	\label{eq:kdotv}
\end{equation}
where $v_{\parallel} \equiv \vvec_0\dotv\hat{\xvec} = v_0\cos\psi$, $v_{\perp} \equiv |\vvec_0\cross\hat{\xvec}| = v_0\sin\psi$, and $\mu \equiv \hat{\kvec}\dotv\hat{\xvec}$.

In these coordinates \eq{eq:response} becomes
\begin{equation}
	\varphi = \frac{q}{(2\pi)^3\varepsilon_0}\int_{-1}^1\!\!\!\! d\mu\!\!\int_0^{2\pi}\!\!\!\!\!\! d\phi\;
	\Re\!\!\int_0^{\infty}\!\!\!\!\! dk\,
	\frac{k^2\exp (ik\mu r)}{k^2 + \Phi(\mu v_{\parallel} + \sqrt{1-\mu^2}\cos\phi\,v_{\perp})} \;,
	\label{eq:phiNum1}
\end{equation}
where $r\equiv |\xvec|$ and $\Phi$ is defined in \eq{eq:Phidef}. The units used were such that $\kDi = \ompi = 1$ and $q = 4\pi\varepsilon_0$.

By using the fact that $\Phi(x)^{*} = \Phi(-x)$ for real $x$, the range of the $\mu$ and $\phi$ integrations was reduced by half. It was found useful\cite{dewar67} to define the new special function $\eta(z)$, \eq{eq:etadef}. Analytical properties of $\eta(z)$ are discussed in Appendix I 
of the thesis\cite{dewar67} and a listing of a Fortran IV subroutine for its efficient evaluation is given. This was used in the code developed to calculate the results presented below. (More discussion of the numerical method is given in the thesis.\cite{dewar67})

In terms of $\eta$, \eq{eq:phiNum1} reduces to a double integration
\begin{equation}
	\varphi = \frac{2}{\pi^2 r}\!\! \int_0^{\pi}\!\!\!\! d\phi\;
	\left\{
	   \frac{\pi}{2} + \int_0^r \!\! dx\,\Im
		\left[ \sqrt{\Phi}\,\eta(x\sqrt{\Phi})
		\right]
	\right\} \;,
	\label{eq:phiNum2}
\end{equation}
where
\begin{equation}
	\sqrt{\Phi} \equiv
	\left\{
		\Phi\left(\frac{xv_{\parallel}}{r} + \left[1 - \left(\frac{x}{r}\right)^2\right]^{1/2}\!\!\!\cos\phi\, v_{\perp}\right)
	\right\}^{1/2} \;.
	\label{eq:sqrtPhidef}
\end{equation}

\subsection{Small $v_0$ results}
\label{sec:Smallv0Num}

\begin{figure}[thb]		
	\centerline{\psfig{file=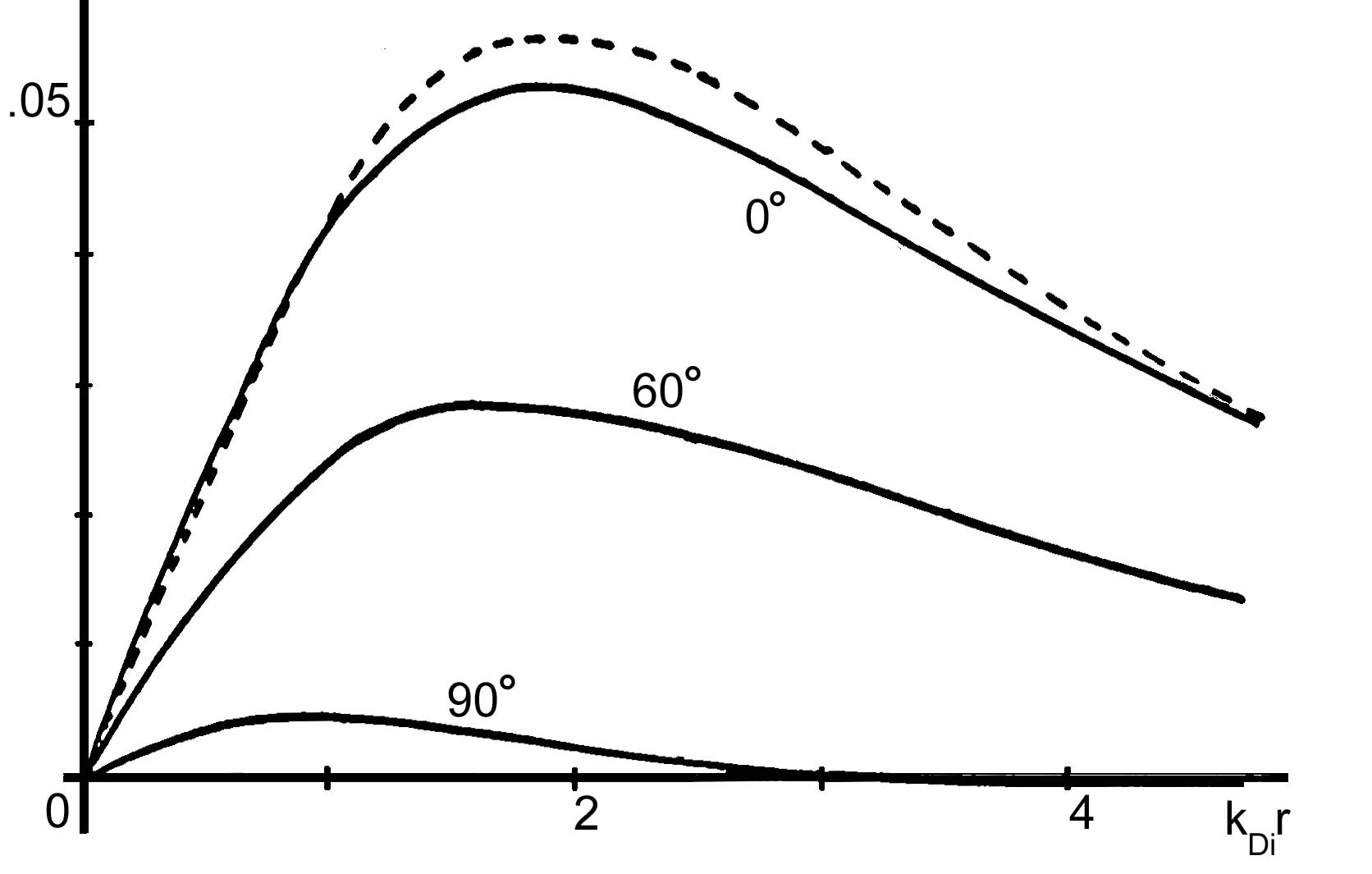,width=4in,angle=0}}
	\vspace*{1pt}
	\caption{Solid curves are the computed deviation from the Debye potential, as defined in the text, for a particle moving at velocity $v_0 = 0.2\ompi/\kDi$ through a plasma with Lorentzian ion distribution and infinite-temperature electrons. Dashed curve is an analytic approximation to the forward field from \eq{eq:varphi2}.
	\label{fig:MScFig3_2}}
\end{figure}

The case where the test particle is travelling slower than the ion thermal speed was studied using the Lorentzian distribution function \eq{eq:gLorentz} with $\kDe = 0$. The treatment of Sec.~\ref{sec:3a_smallv0} indicates that the choice of distribution function and electron Debye length simply alters the relative magnitude of the constants $k_{\rm D}$, $a_0$ and $a_1$, so we expect there is little to be gained by study of a large variety of cases.

\begin{figure}[thb]		
	\centerline{\psfig{file=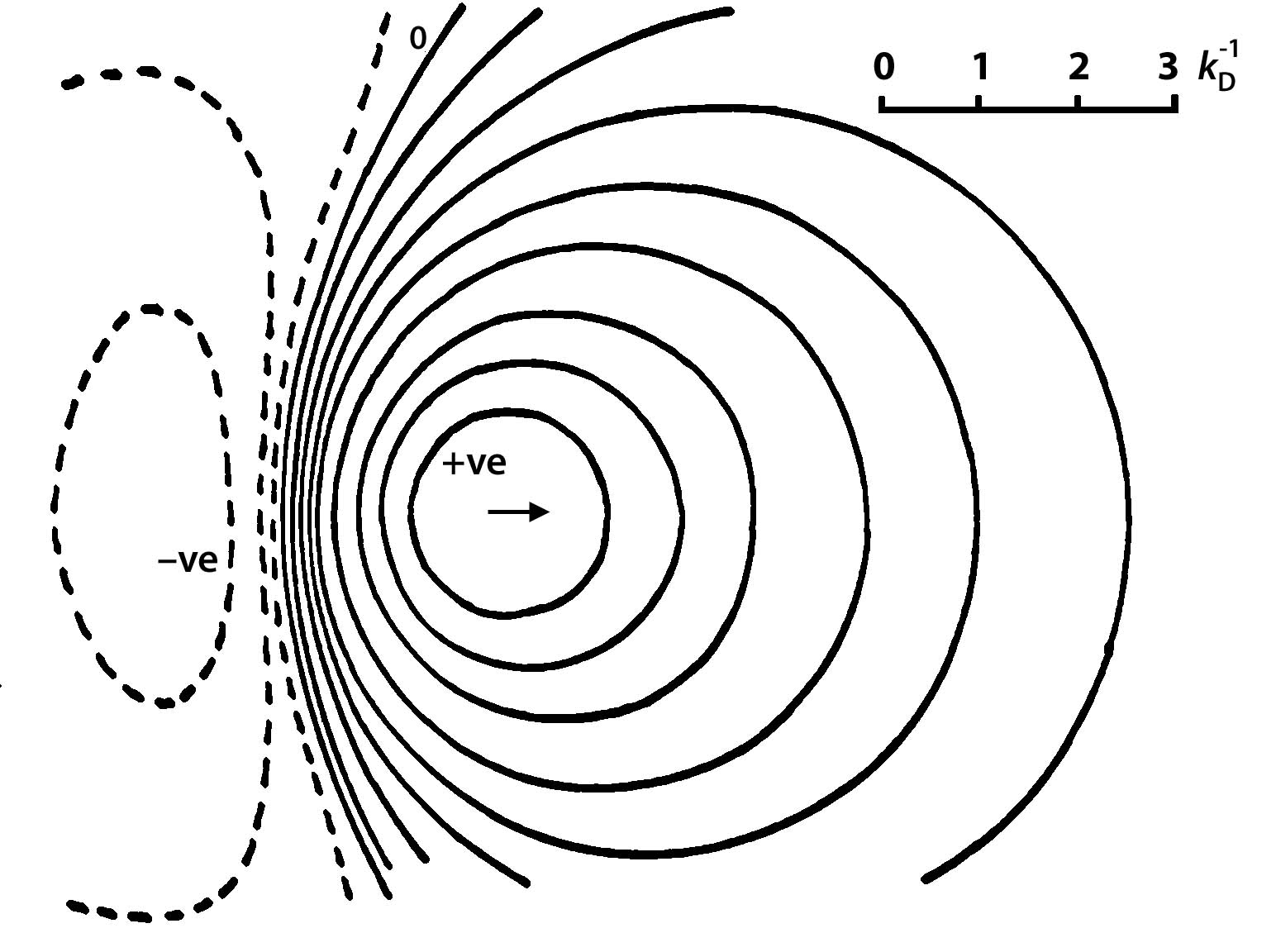,width=4in,angle=0}}
	\vspace*{1pt}
	\caption{Equipotentials in a plasma the same as assumed for \fig{fig:MScFig3_2}  but with $v_0 = 0.4\ompi/\kDi$ (the direction of $\vvec_0$ is indicated by an arrow). The solid equipotentials indicate positive ($+$ve) $\varphi$ and the dashed lines negative ($-$ve). The scale in Debye lengths is indicated top right.
	\label{fig:MScFig3_3}}
\end{figure}

In \fig{fig:MScFig3_2} we plot the deviation from the Debye potential along lines making angles $\psi$ of $0^{\circ}$, $60^{\circ}$ and $90^{\circ}$ to the direction of travel, the deviation being defined as
$[\varphi(\xvec) - \varphi_{\rm D}(r)]/\varphi_0(r)$, where $\varphi_{\rm D}(r) \equiv q \exp(-\kDi r)/4\pi\varepsilon_0 r$ is the Debye potential and $\varphi_0(r) \equiv q/4\pi\varepsilon_0 r$ is the bare, unscreened potential. The approximate result at $0^{\circ}$ obtained from the term of \eq{eq:varphi2} first order in $v_0$ is also plotted (dashed curve), the agreement being seen to be quite reasonable.

The qualitative form of \fig{fig:MScFig3_3} is also in agreement with that predicted from the analytical work. Note the potential well behind the particle, beyond which the field becomes attractive to charges of the same sign as the test particle.


\subsection{Intermediate $v_0$ results}
\label{sec:Intermediatev0Num}

\begin{figure}[thb]		
	\centerline{\psfig{file=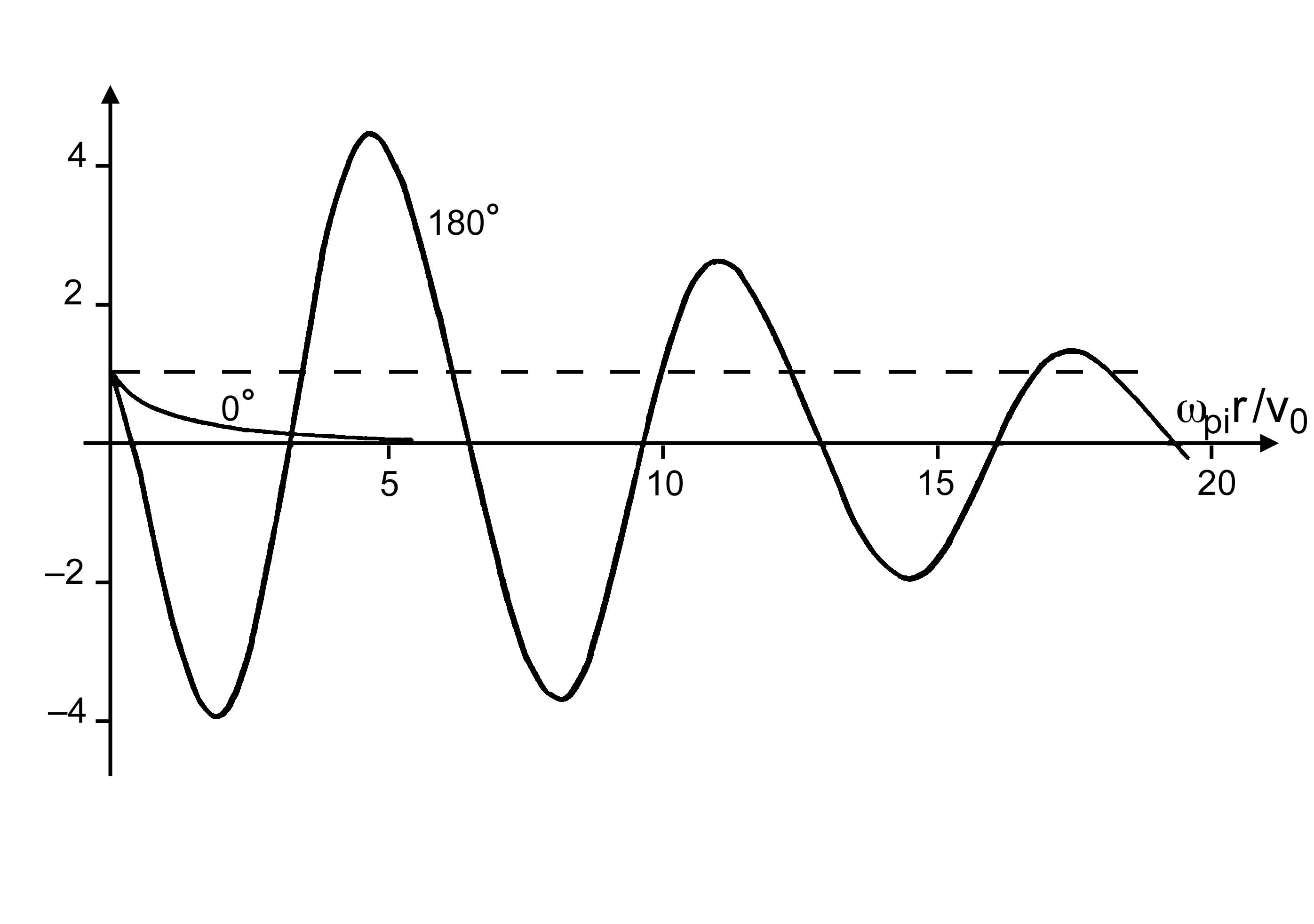,width=3.5in,angle=0}}
	\vspace*{-24pt}
	\caption{Plots of $\varphi(r,\psi)/\varphi_0(r)$ as a function of distance $r$ (in units of $v_0/\ompi$) calculated for a test particle moving at $v_0 = 8\ompi/\kDi$ in a plasma with Lorentzian ions and $\kDe = 0$. The forward field is labelled $0^{\circ}$ and the wake field is labelled $180^{\circ}$. The dashed line is the unscreened case, $\varphi = \varphi_0$.
	\label{fig:MScFig3_4}}
\end{figure}
\begin{figure}[thb]		
	\centerline{\psfig{file=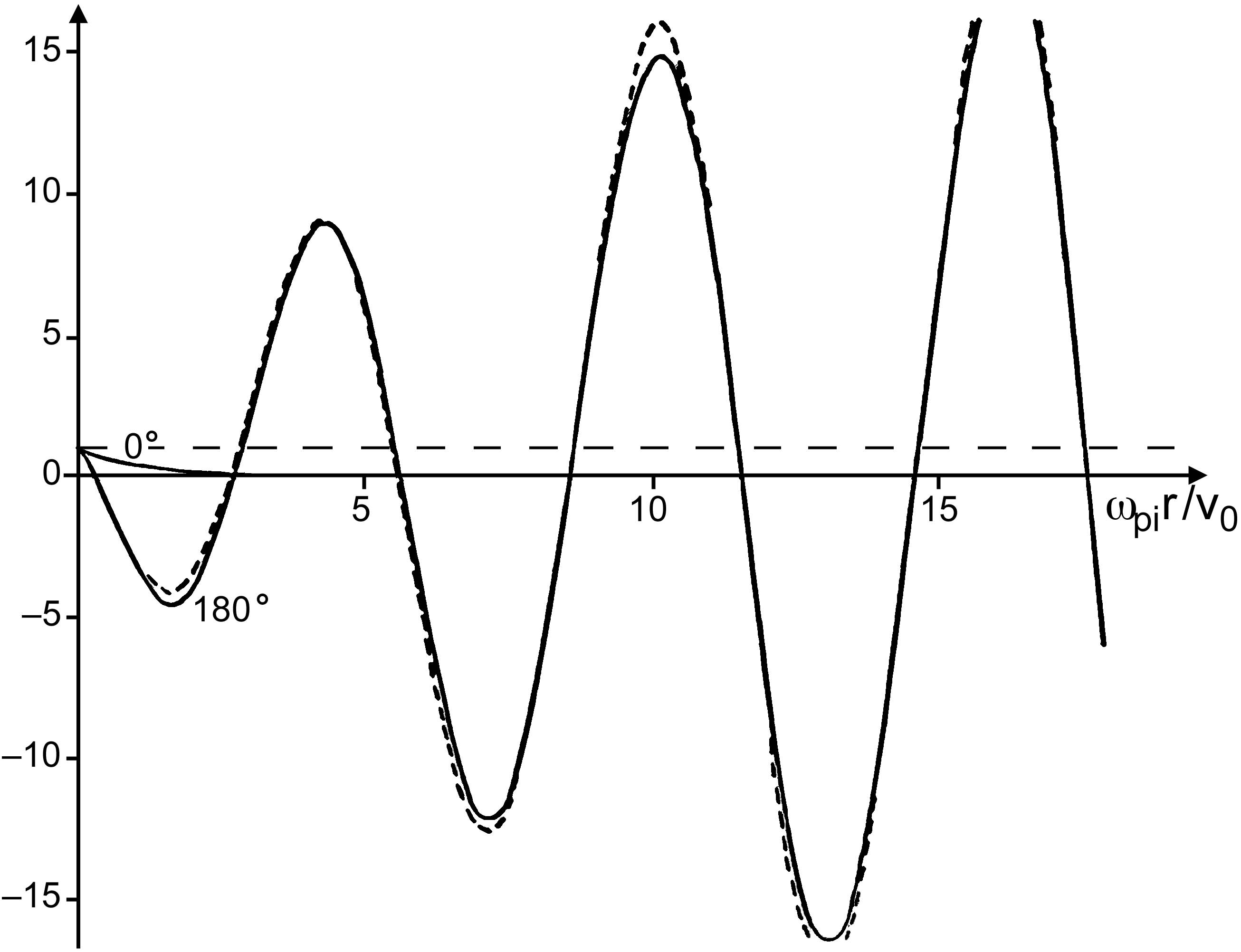,width=3.5in,angle=0}}
	\vspace*{-6pt}
	\caption{Plots of $\varphi(r,\psi)/\varphi_0(r)$ as a function of distance $r$ (in units of $v_0/\ompi$) calculated for a test particle moving at $v_0 = 8\ompi/\kDi$ in a plasma with Maxwellian ions and $\kDe = 0$. The forward field is labelled $0^{\circ}$ and the wake field is labelled $180^{\circ}$. The short-dashed curve is the approximation \eq{eq:Ib0}.
	\label{fig:MScFig3_5}}
\end{figure}

The intermediate case, where the test particle moves faster than the ion thermal speed but less than the ion sound speed  $C_{\rm s}$, was studied in part using both Lorentzian and Maxwellian distribution functions as these to some extent represent extremes of Landau damping. The Lorentzian distribution is sufficiently simple for an exact dispersion relation to be derivable, thus obviating the use of asymptotic expansions of $\Phi$. Again we took $\kDe = 0$, so  $C_{\rm s} = \infty$ by \eq{eq:vpmax}.

\begin{figure}[thb]		
	\centerline{\psfig{file=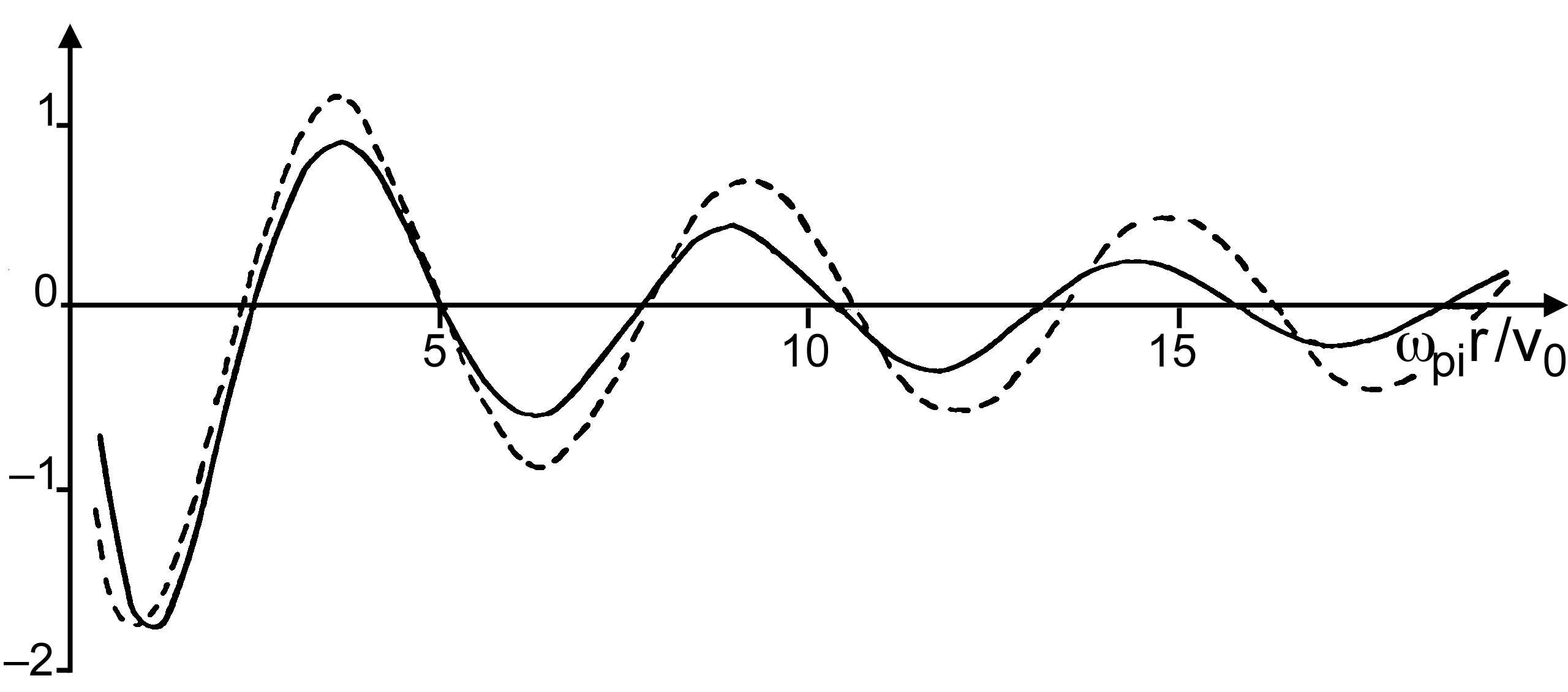,width=3.5in,angle=0}}
	\vspace*{-6pt}
	\caption{Same case as in \fig{fig:MScFig3_5} except that the test particle speed is halved, $v_0 = 4\ompi/\kDi$. Plotted is $[\varphi(r,\psi)/\varphi_0(r)](\ompi r/v_0)^{-1}$ in the wake, $\psi = \pi$.
	\label{fig:MScFig3_6}}
\end{figure}

Potentials on the axis of symmetry, parallel (forward field, $\psi = 0$) or antiparallel (wake field, $\psi = \pi$) to the direction of motion of the test particle, are plotted in Figs.~\ref{fig:MScFig3_4}--\ref{fig:MScFig3_6}:

\begin{figure}[thb]		
	\centerline{\psfig{file=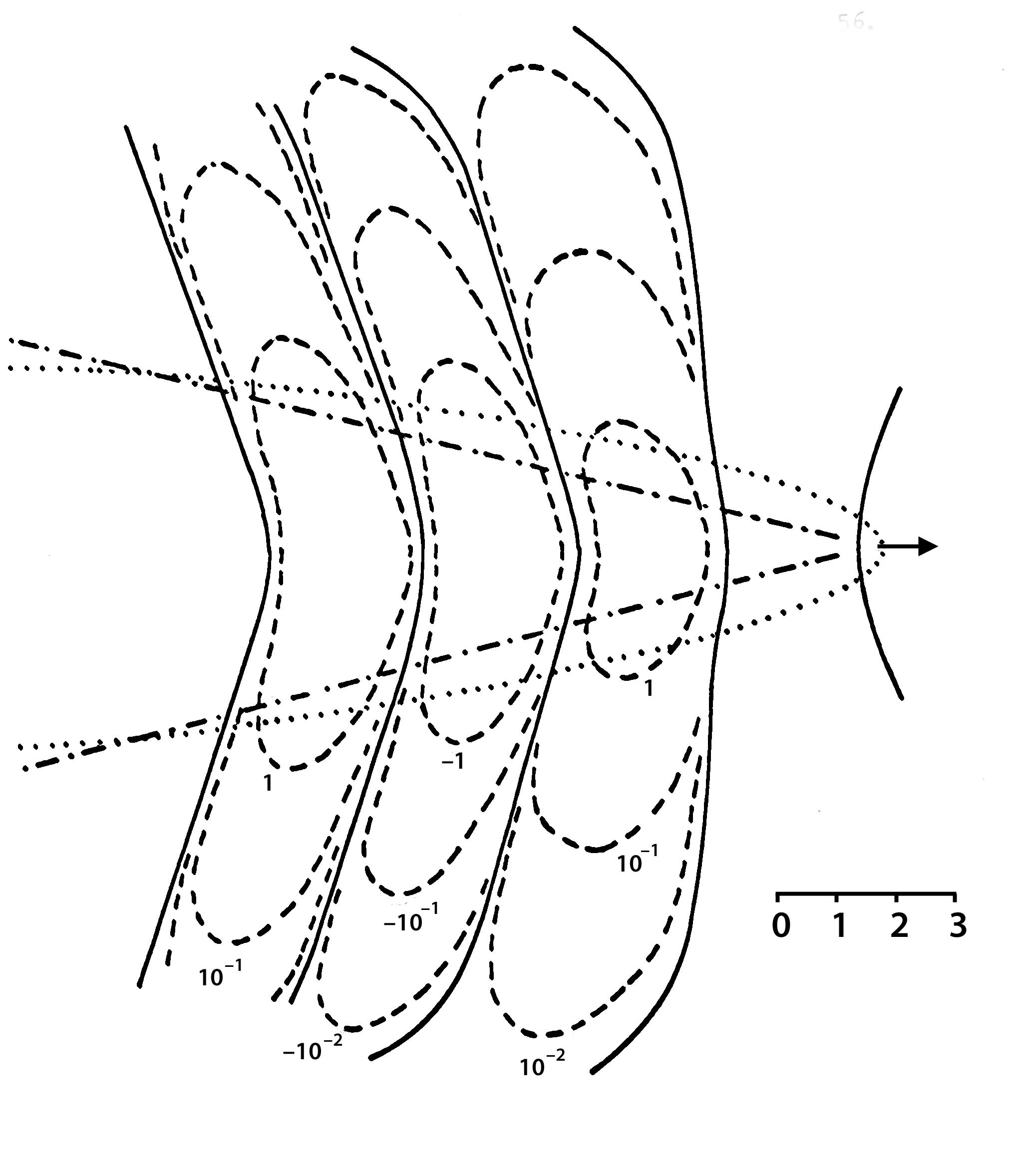,width=3.5in,angle=0}}
	\vspace*{-18pt}
	\caption{Contours of $\varphi(\xvec)/\varphi_0(r)$ for a test particle moving with $v_0 = 4\ompi/\kDi$ (the direction of $\vvec_0$ is indicated by an arrow) in a plasma with Maxwellian ions and $\kDe = 0$. The scale in units of $v_0/\ompi$ is indicated lower right, the ``thermal Mach cone'' $x_{\perp} = |x_1|\ompi/\kDi v_0$ is indicated by dash-dotted lines, and the boundary of the region of validity of \eq{eq:integralsaddle2} is indicated by the dotted curve (cf. \fig{fig:MScFig2_6}).
	\label{fig:MScFig3_7}}
\end{figure}

\noindent\emph{Forward field:} In all three of the plots the results agreed well with \eq{eq:screenedpotlxp0}, with agreement particularly good for the Maxwellian case (four significant figures at $v_0 = 8\ompi/\kDi$. [The potentials at $90^{\circ}$ to the direction of motion also agreed reasonably with \eq{eq:screenedpotlx10}.]

\noindent\emph{Wake field, Lorentzian case, \fig{fig:MScFig3_4}:} Within the wake Landau damping had a dominant effect in the Lorentzian case even at the centre of the wake. This is due to the slow decay of the tail of the distribution function. The analytical prediction (which was not derived in Sec.~\ref{sec:3_Analapprox}) was virtually indistinguishable on the scale of the graph, which is to be expected because the dispersion relation is exact.

\begin{figure}[thb]		
	\centerline{\psfig{file=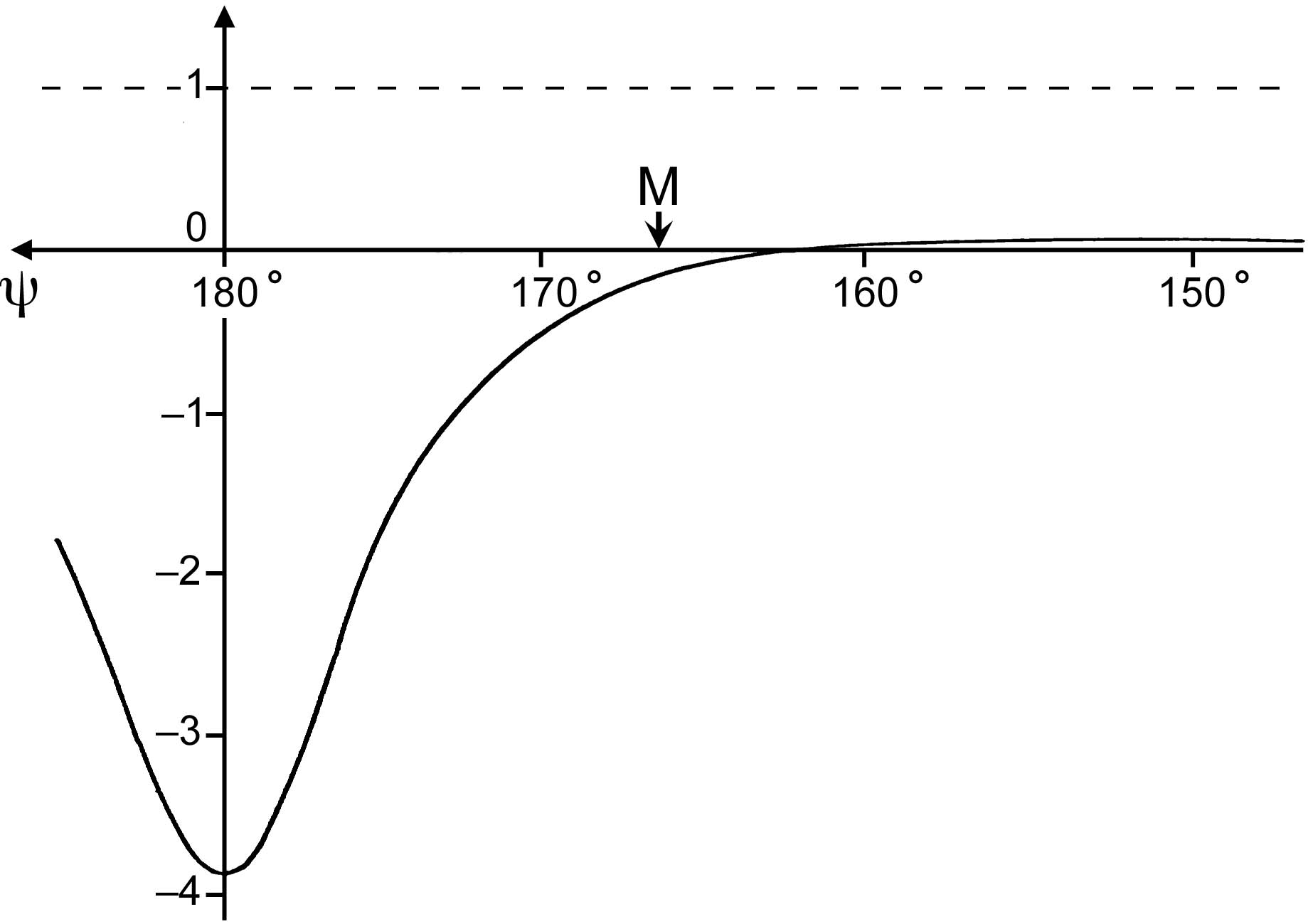,width=3.5in,angle=0}}
	\caption{Plot of $\varphi(r,\psi)/\varphi_0(r)$ \emph{vs.} $\psi$ (with $r = 48/\kDi$) in the wake of a supersonic test particle moving at $v_0 = 32\ompi/\kDi = 4C_{\rm s}$ in a plasma with $\kDe = \kDi/8$. Surprisingly, no shock wave is seen at the expected position of the Mach cone, which is calculated from $x_{\perp}/|x_1| = \ompi/v_0\kDe$ and indicated by the arrow marked M at $\psi = 166^{\circ}$.
	\label{fig:MScFig3_8}}
\end{figure}

\noindent\emph{Wake field, Maxwellian case:} For a Maxwellian distribution function the agreement with \eq{eq:Ib0} was quite good in the $v_0 = 8$ case, \fig{fig:MScFig3_5}, but not so good in the $v_0 = 4$ case, \fig{fig:MScFig3_6}. This may be explained by the inadequacy of the asymptotic expansion \eq{eq:PhiAsymp} when $|\omega/k| < 4$. Landau damping may also have some effect.

Owing to computing time limitations the complete wake structure was mapped only in the $v_0 = 4$ Maxwellian case. A contour plot is given in \fig{fig:MScFig3_7}, which represented almost two hours of computing time on the IBM 7044 and is still not very accurate.

\subsection{Supersonic case}
\label{sec:Supersonicv0Num}

The supersonic case, $v_0 > C_{\rm s}$, considered in Sec.~\ref{sec:3cii_supersonicv0} was only briefly studied numerically. Figure~\ref{fig:MScFig3_8} shows a fairly sharp cutoff near the Mach cone but no shock front. Indeed the function decays monotonically off the axis in complete contrast with the Kraus--Watson\cite{kraus-watson58} result, which predicts an initial increase.

\subsection{Conclusion}
\label{sec:Concl}

In the intervening years since my MSc work other authors have made similar calculations. I have not attempted a complete literature search, but note that the research project must have been topical at the time as the inverse third power asymptotic behaviour in \eq{eq:varphi3}  was announced a year later by Montgomery, Joyce and Sugihara,\cite{Montgomery_Joyce_Sugihara_1968} in a paper that has been cited 58 times. In fact the problem is even more topical today with the rise of interest in dusty plasmas---for instance the paper by Ishihara and Vladimirov\cite{Ishihara_Vladimirov_1997} on the wake potential of a dust grain in a plasma with ion flow has been cited more than 80 times.

\section*{Acknowledgments}
I am grateful to Ken Hines for providing an ambiance in which we research students were able to develop intellectually both through his gentle guidance and through mutual interactions. I am indebted particularly to Norm Frankel for indoctrinating me in statistical physics and kinetic theory and sharing his thoughts on many topics, and to Andrew Prentice for many stimulating conversations and for providing the subroutine I adapted for calculating the response function $\Phi$.

\bibliography{TestParticle,RLDBibDeskPapersX}

\end{document}